\documentclass[aps,pra,superscriptaddress,twocolumn,showpacs,noeprint,notitlepage,longbibliography]{revtex4-2}

\usepackage{graphicx}% Include figure files
\usepackage{dcolumn}% Align table columns on decimal point
\usepackage{bm}% bold math
\usepackage{tikz-cd}
\usepackage{mathrsfs}
\usepackage{algpseudocode}
\usepackage[caption=false]{subfig}
\usepackage{graphicx,amsfonts,times,bm,amsmath,amssymb,verbatim,ifthen,braket,xcolor,array,natbib,comment,enumerate,multirow,soul,lineno,algorithm,algpseudocode,hyperref}
\usepackage{ulem}

\begin{document}

\title{Generalized Pulse Design in Floquet Engineering: Application to Interacting Spin Systems}

\author{Ryan Scott}
\affiliation{Department of Physics, Virginia Tech, Blacksburg, Virginia 24061, USA}

\author{Bryce Gadway}
\affiliation{Department of Physics, The Pennsylvania State University, University Park, Pennsylvania 16802, USA}

\author{V. W. Scarola}
\email[Email address:]{scarola@vt.edu}
\affiliation{Department of Physics, Virginia Tech, Blacksburg, Virginia 24061, USA}

\begin{abstract}
Floquet engineering in quantum simulation employs externally applied high frequency pulses to dynamically design steady state effective Hamiltonians.  Such protocols can be used to enlarge the space of Hamiltonians but approximations often limit pulse profile shapes and therefore the space of available effective Hamiltonians.  We consider a non-stroboscopic high frequency expansion formalism for Floquet engineering.  We generalize the pulse profiles available by rigorously keeping all necessary terms to lowest order in inverse frequency expansions used to derive the effective Hamiltonians.  Our approach allows wide tunability in application of external driving fields.  We apply our method to long-range interacting XXZ spin Hamiltonians.  We model an example application where we derive conditions on specific pulse shapes to engineer effective Ising models from XXZ models.  Our method allows the space of continuous pulse profiles, relevant to experimental control fields, to better and more accurately explore possible effective Hamiltonians available for Floquet engineering. 
\end{abstract}

\maketitle

\section{Introduction}

Broadening the class of accessible quantum simulation models by improving the precision and applicability of engineering protocols will help experiments study models otherwise inaccessible to conventional numerical methods \cite{BLOCH2012,BLATT2012,Georgescu2014,Gross2017,MONROE2021,MORGADO2021a,DALEY2022}.  High frequency driving (Floquet engineering) with external fields has proven to be a useful experimental tool whereby otherwise static system Hamiltonians, $\hat{H}_0$, are perturbed with periodic external field pulses, $\hat{V}(t)$.  The resulting driven Hamiltonian:
\begin{equation}
    \hat{H}(t) = \hat{H}_0 + \hat{V}(t),
    \label{eq:hamiltonianSystem}
\end{equation}
can, with the appropriate pulse design, lead to a desired averaged or effective Hamiltonian, $\hat{H}_{\text{eff}}$, that is qualitatively distinct from $\hat{H}_0$.  Methods to derive $\hat{H}_{\text{eff}}$ often use perturbative expansions in the high frequency limit, where the driving frequency is:
\begin{align}
\omega=\frac{2\pi}{T},
\end{align}  
and $T$ is a subcycle time defining periodicity such that $\hat{V}(t)$ is periodic in an integer multiple of $T$.  $\omega^{-1}$ establishes a small expansion parameter compared to all other energy scales in $\hat{H}_0$. Interesting properties of eigenstates of $\hat{H}_{\text{eff}}$ can then be observed experimentally.    

Recent experiments with Rydberg atoms \cite{BLUVSTEIN2021a,GEIER2021c,scholl2022microwave,ZHAO2023}, molecules \cite{CHRISTAKIS2023b}, and trapped ions \cite{MORONG2023} have been able to dynamically engineer effective higher symmetry Heisenberg spin models from lower symmetry models.  Ref.~\onlinecite{GEIER2021c} used rectangular pulse profiles to engineer an effective $SU(2)$-symmetric Heisenberg interaction between a large number of Rydberg atoms initially lacking this symmetry in their interactions.  Similarly, trapped ions were recently driven \cite{MORONG2023} from an Ising limit into a higher-symmetry Heisenberg regime to replicate the Haldane-Shastry model \cite{HALDANE1988b,SHASTRY1988}.  Interest in engineering systems extends beyond transformations of underlying symmetry groups, or simulating $SU(2)$ Hamiltonians. For example, there have been proposals and experiments employing cold-atom systems to explore gauge theories and other field theories \cite{halimeh2025cold,ott2021scalable,sundar2018synthetic,stannigel2014constrained}, and trapped ions have been used to simulate complex open systems  \cite{barreiro2011open,MONROE2021}. 

The theory of Floquet engineering, and the associated high frequency expansions, has a long history (see Refs.~\cite{BLANES2009a,LESKES2010,goldman2014periodically,bukov2015universal} for reviews). There are two common methods to perform controlled expansions in inverse frequency that allow for effective engineering of broad classes of Hamiltonians. The first is the high frequency stroboscopic Magnus expansion (SME) \cite{MAGNUS1954,bukov2015universal}. SMEs are commonly used in derivations of $\hat{H}_{\text{eff}}$, particularly in the context of average Hamiltonian theory \cite{HAEBERLEN1968b,LESKES2010}. SMEs can be pushed to high order in inverse frequency to derive non-trivial effective models, e.g., multi-spin interactions \cite{HERDMAN2010,choi2020robust}, but often layer in additional assumptions. The second is the gauge invariant high frequency expansion (GIHFE) \cite{rahav2003effective,goldman2014periodically,bukov2015universal,Abanin2017}. The GIHFE approach relies on the definition of a kick operator, $\hat{K}(t)$, from $\hat{V}(t)$ \cite{goldman2014periodically}.  The effective Hamiltonian and new gauge transformed state follow from:
\begin{align}
        \hat{H} &\rightarrow \hat{H}_{\text{eff}} = e^{i \hat{K}(t)}\hat{H}e^{-i \hat{K}(t)} + i \left(\frac{\partial e^{i\hat{K}(t)}}{\partial t}\right)e^{-i\hat{K}(t)},\nonumber \\
        \psi(t) &\rightarrow \phi(t) = e^{i \hat{K}(t)} \psi(t).
        \label{eq:gaugeTransformation}
\end{align}
From these expressions we see that derivation of $\hat{H}_{\text{eff}}$ and derivation of the new states $\phi(t)$ from the original $\psi(t)$ then requires a systematic inverse frequency expansion of $\hat{K}(t)$ \cite{rahav2003effective,goldman2014periodically,bukov2015universal} in addition to the expansion for $\hat{H}_{\text{eff}}$.  Using $\hat{H}_{\text{eff}}$, $\hat{K}(t)$, and $\phi(t)$ one can then compute dressed (time averaged) observables \cite{bukov2015universal}.

GIHFEs, in contrast to SMEs, are often used to model sampling even at non-stroboscopic times by tracking the ensuing change in basis (gauge transformation) initiated by pulsing \cite{Bukov2014}. Despite the distinction, one can recover average Hamiltonian theory from GIHFEs by restricting to stroboscopic sampling and applying the usual Magnus expansion \cite{bukov2015universal}. The non-stroboscopic case can be constructed via a gauge transformation that depends on the initial sampling time.  Assumptions underlying conventional GIHFE methods limit the space of available pulse profiles and therefore limit possible effective Hamiltonians.  

We revisit the GIHFE formalism with the aim of broadening the set of available effective Hamiltonians in driven spin systems.  The engineering method we construct generalizes the available pulse shapes considered in the existing literature \cite{rahav2003effective,goldman2014periodically,bukov2015universal}.  We consider periodic pulse structures such that $\omega^{-1}$ defines the \textit{only} small parameter allowing  otherwise arbitrary pulse shapes.  Our formalism then allows for large pulse strengths, any pulse width that respects the high frequency approximation, and non-stroboscopic sampling.   The pulse shape freedom we find arises because we are able to leverage the commutation relations of the Pauli spin matrices to perform controlled series expansions approximating $\hat{K}(t)$.   Hamiltonian engineering then reduces to a functional optimized over the entire space of pulse shapes so that the desired effective Hamiltonian is realized. 

To exemplify our approach, we apply our formalism to \textit{decrease} the symmetry implied by native spin-spin interactions, motivated by the desire to construct (weighted) graph states \cite{Hein2006a} in spin systems with native dipolar XXZ spin-spin interactions which characterize certain Rydberg atom \cite{Saffman2010,NGUYEN2018} and molecule \cite{Barnett2006,WALL2010a,muller2010,Wall2014} interactions. Since XXZ interactions do not commute in general, the protocol required to build weighted graph states introduces complexities that the Ising-type interaction avoids \cite{Hein2006a,tanamoto2009efficient}. Requirements for building high precision weighted graph states with global pulses therefore motivates the specific example we use to demonstrate our formalism: engineering an XXZ interaction into an Ising interaction. 
 
 The paper is organized as follows. In Sec.~\ref{sec:spinmodels} we define a class of long-range interacting spin models with various limits of interest.  In Sec.~\ref{sec:Formalism} we present our formalism where Sec.~\ref{sec:Assumptions} introduces our core set of assumptions.  Sec.~\ref{sec:strongdriving} then uses these assumptions to derive a Floquet engineering formalism applicable to pulses with arbitrary shape for use at non-stroboscopic times.  Sec.~\ref{sec:xxmodel} applies our formalism to XXZ-type interacting spin systems.  As a demonstration, we define specific global cosine and square pulses and establish the conditions under which an XXZ model can be accurately engineered into an Ising model.  We summarize in Sec.~\ref{sec:summary}.

\section{Spin Models}
\label{sec:spinmodels}

We consider the following example spin models for $\hat{H}_0$, setting $\hbar = 1$ throughout:
\begin{equation}
    \hat{H}^{\text{S}}_0 = \frac{1}{2}\sum_{j \neq j'}\left[J^1_{j j'}  \hat{\sigma}^{(1)}_{j}\hat{\sigma}^{(1)}_{j'} + J^2_{j j'}\hat{\sigma}^{(2)}_{j} \hat{\sigma}^{(2)}_{j'} + J^3_{j j'}\hat{\sigma}^{(3)}_{j}\hat{\sigma}^{(3)}_{j'} \right],
    \label{eq:spinModels}
\end{equation}
where $\hat{\sigma}^{(1)}_{j}$, $\hat{\sigma}^{(2)}_{j}$, and $\hat{\sigma}^{(3)}_{j}$ denote the usual $x$, $y$, and $z$ Pauli-spin matrices, respectively.  $j$ and $j'$ denote site (vertex) pairs in an arbitrary graph.  $J^1_{j j'}$, $J^2_{j j'}$, and $J^3_{j j'}$ denote spatially varying spin-spin interaction energies along the spin-$x$,$y$, and $z$ directions, respectively.  Our formalism applies to more general models, but we will use $H_0$ as an example for applications.  In specifying example applications, we restrict our attention to XXZ-type models with:
\begin{equation}
    \begin{split}
        J^1_{j j'} &= J^2_{j j'} =J_{\perp} \mathcal{V}_{j,j'} \\
        J^3_{j j'} &= J_z \mathcal{V}_{j,j'},
    \end{split}
    \label{eq:Jsdipolar}
\end{equation}
where $\mathcal{V}_{j,j'}$ is a dimensionless spin-spin interaction, e.g., the dipolar interaction is retrieved when 
$\mathcal{V}_{j,j'}$ is chosen to be $|j-j'|^{-3}$.  $J_{\perp}$ and $J_{z}$ denote interaction energy constants. 

Equation~\eqref{eq:spinModels} encompasses well-known spin models.  For $J_z=0$, Eq.~\eqref{eq:spinModels} reduces to an XY-model, denoted $H_{XY}$. For $J_{\perp}= 0$, we retrieve the Ising model, denoted $H_{ZZ}$.  Eq.~\eqref{eq:Jsdipolar} implies that we may write $\hat{H}_0^{\text{S}}$ as $\hat{H}_{XXZ} = \hat{H}_{XY} + \hat{H}_{ZZ}$. 
Other choices of the interaction constants give rise to additional models, such as the Heisenberg model ($J_{\perp}=J_z$).  Our formalism will hold for general spin systems not necessarily conforming to Eq.~\eqref{eq:Jsdipolar}, but we will consider a specific example of Hamiltonian engineering below where we seek to engineer $\hat{H}_{XXZ}$ into $\hat{H}_{ZZ}$. Although we restrict to general 2-body interactions as given by Eq.~\eqref{eq:spinModels}, our considerations here can be generalized in a straightforward manner to higher order interactions.

We motivate the need for a Floquet formalism that allows arbitrary pulse structures with an example.  Consider engineering of the 2-site XXZ model:
\begin{equation}
    \hat{H}_0^{\text{S},2} = J_{\perp}(\hat{\sigma}^{(1)}_1\hat{\sigma}^{(1)}_2 + \hat{\sigma}^{(2)}_1\hat{\sigma}^{(2)}_2) + J_z \hat{\sigma}^{(3)}_1\hat{\sigma}^{(3)}_2,
\end{equation}
into an Ising model.  Driving one of the spins in the XXZ interaction in the $xy$-plane, i.e., rotation about the $z$-axis, should cancel the XY-interaction term.  To this end, consider a periodic driving pulse, $\hat{V}(t)$, of a single frequency  given by  $V_0 \hat{\sigma}^{(3)}_2 \cos(\omega t)$.  We can use the Jacobi-Anger relation to show that it is possible drive $\hat{H}_0^{\text{S},2}$ into $\hat{H}_{ZZ}$ with $\vert V_0 \vert\sim\omega$ (See, e.g., Ref.~\onlinecite{SOOMRO2022}, for a related derivation).  However, rigorous application of the Jacobi-Anger method is limited to a single frequency cosine driving pulse and application to general pulse profiles is not straightforward. 

To engineer $\hat{H}_0^{\text{S},2}$ into an Ising model using a more general method (applicable to other pulse shapes), we can attempt to use the formalism of Ref.~\onlinecite{goldman2014periodically}. This formalism can apply to pulse shapes of many frequencies, however the main results of Ref.~\onlinecite{goldman2014periodically} only apply when the pulse is weak, $\vert V_0 \vert \ll \omega$.  To see this, we use Ref.~\onlinecite{goldman2014periodically} by taking Fourier coefficients of $\hat{V}(t)$, and expand to lowest order in the effective Hamiltonian to arrive at a static effective model: 
\begin{equation}
    \hat{H}_{\text{eff}}^{\text{S},2} = \hat{H}_0^{\text{S},2}  - \frac{V_0^2}{4 \omega^2}[\hat{\sigma}^{(3)}_2, [\hat{\sigma}^{(3)}_2, \hat{H}_0^{\text{S},2}]] + \mathcal{O}(\omega^{-3}).
    \label{eq:2BodyEffective}
\end{equation}
Evaluating commutators for the 2-site XXZ model yields:
\begin{align}
        \hat{H}_{\text{eff}}^{\text{S},2} &= J_{\perp} \left(1-  \frac{V^2_0 }{\omega^2} \right)(\hat{\sigma}^{(1)}_1\hat{\sigma}^{(1)}_2 + \hat{\sigma}^{(2)}_1\hat{\sigma}^{(2)}_2) + J_z \hat{\sigma}^{(3)}_1\hat{\sigma}^{(3)}_2 \nonumber,
\end{align}
where the periodic driving leads to a perturbation of strength $V^2_0/\omega^2$ that effectively subtracts an XY contribution, as intended. We therefore see that the family of effective models given by $\hat{H}_{\text{eff}}^{\text{S},2}$ as we vary the pulse structure is limited; in particular, $\hat{H}_{\text{eff}}^{\text{S},2}$ was derived under the assumption $\vert V_0 \vert \ll \omega$.  If we seek to engineer an effective Ising model by forcing the XY terms to cancel: 
\begin{equation}
 \hat{H}_{\text{eff}}^{\text{S},2} \rightarrow J_z \hat{\sigma}^{(3)}_1\hat{\sigma}^{(3)}_2,
\end{equation} 
we must require $ \omega = |V_0|$, which violates the condition $\vert V_0 \vert \ll \omega$.  Similar issues arise in attempting pulses with other spatial profiles, e.g., experimentally relevant global pulses applied simultaneously to all sites.  Using the main results of  Ref.~\onlinecite{goldman2014periodically} thus requires a generalization to incorporate arbitrarily strong pulse profiles.

The correction term appearing in Eq.~\eqref{eq:2BodyEffective} arises from a stroboscopic approximation to the GIHFE.  This approximation is equivalent to an SME. The underlying assumptions made to perform the calculation are common, but give rise to limitations that prevent more general engineering than the calculation in Eq.~\eqref{eq:2BodyEffective} suggests. 

\begin{figure}[t]
    \includegraphics[width=0.95\linewidth]{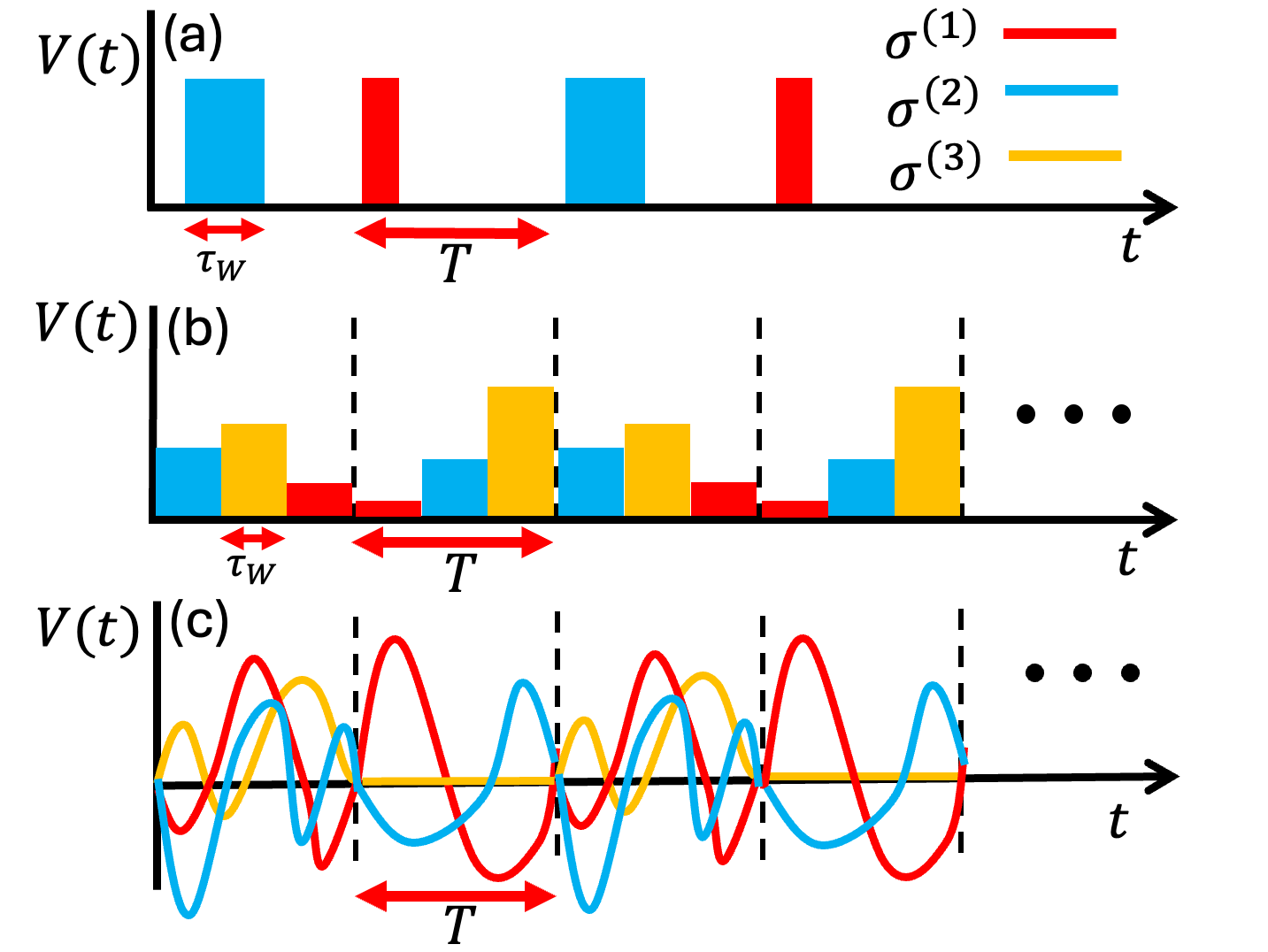}
    \caption{Schematic time-dependence of example pulse profiles, $\hat{V}(t)$ with cycle time $2T$. Colored bars and lines indicate pulsed application of Pauli operators, specifically application of $h_{j}^{\alpha}(t) \hat{\sigma}_{j}^{(\alpha)}$ in Eq.~\eqref{eq:perturbingPotential} such that  $\hat{\sigma}^{(1)},\hat{\sigma}^{(2)},$ and $\hat{\sigma}^{(3)}$ denote the usual Pauli $x,y,$ and $z$ matrices, respectively.  $T$ indicates the duration of a repeated pulse subcycle. (a) A conventional rectangular pulse profile where $\tau_w$ labels the pulse width.  (b) The same as (a) but with pulses designed to replicate a continuous pulse profile with discrete pulses.  (c) A general continuous pulse profile that can be accurately implemented with our approach.}
    \label{fig:PulseGeneralization}
\end{figure}

As an additional example, consider the commonly applied global rectangular profiles. Fig.~\ref{fig:PulseGeneralization}a shows an illustrative example.  While the periodic driving is assumed to be fast with respect to other time scales set by $\hat{H}_0$, the pulse itself introduces new times scales and other assumptions: i) Precision quantum simulation experiments using smooth approximations to rectangular pulses often need to account for the discrepancy between ideal and realized pulses--see, e.g., Ref. \onlinecite{MORONG2023};  ii)   The pulse strength introduces an energy scale.  The simplest approximations assume that the pulse strength is much smaller than $\omega$,  but this assumption limits tunability of  $\hat{H}_{\text{eff}}$ unless special assumptions are made; iii) The pulse width, $\tau_w$, must be assumed to be much shorter than $T$ but is bounded below by experimental constraints.   For example, practical constraints on the $\tau_w\ll T$ assumption were studied in Ref.~\onlinecite{GEIER2021c} using SMEs to show that finite pulse width led to an error in $\hat{H}_{\text{eff}}$ that scales as $\sim \tau_w/T$ with an impact on measurements of magnetization in their spin system to be as large as $10\%$.  

These examples highlight the need for a rigorous \textit{general} formalism with minimal assumptions on pulse shapes and controlled error. A step toward the generalization was considered in Appendix K of Ref.~\onlinecite{goldman2014periodically}, where a single non-stroboscopic (global) pulse was considered with an arbitrary amplitude.  Our central aim is to advance the approach discussed in Ref.~\onlinecite{goldman2014periodically} by constructing a formalism for Floquet engineering of non-stroboscopic effective Hamiltonians with a minimal set of assumptions on pulse shape and error controlled by only one small  parameter, $\omega^{-1}$. We relax the assumption that the driving potential appears as a single pulse, and turn to the exposition of the formalism now.

\section{Formalism} 
\label{sec:Formalism}

We implement an effective Hamiltonian formalism \cite{goldman2014periodically} by assuming a time-independent Hamiltonian, $\hat{H}_0$, with a periodic time-dependent perturbing potential, $\hat{V}(t)$. The formalism allows the transformation to a frame that yields a time-independent effective description enabled by expanding in powers of $\omega^{-1}$.  The lowest order expansion yields effective Hamiltonians, $\hat{H}_{\text{\text{eff}}}$, and kick operators derived from a broad class of perturbing potentials  without introducing additional time scales, imposing stroboscopic restrictions, or additional perturbative truncations. First, we will discuss the key assumptions defining the formalism.  We then turn to the explicit theoretical description. 

\subsection{Assumptions} 
\label{sec:Assumptions}

Throughout this work, we make the following assumptions:
\begin{enumerate}
    \item The perturbing potential is cyclic and periodic.
    \item All observables are spanned by tensor products of generators of $SU(2)$, the usual Pauli spin operators, as well as the identity. In particular, the perturbing potential will assumed to be linear in the generators of $SU(2)$.
    \item The lowest order kick operator will be assumed to be the time integral of the perturbing potential. As a special case, the perturbing potential will represent global, time-ordered pulses which do not overlap in time.
    \item The effective Hamiltonian and kick operators can be expanded in powers of $\omega^{-1}$ such that the expansion converges for times of interest.
\end{enumerate}
The next four sections discuss these assumptions. 

\subsubsection{Cyclic, Periodic Perturbing Potential} \label{sec:CyclicPeriodic}

As is common \cite{HAEBERLEN1968b,LESKES2010,bukov2015universal,goldman2014periodically}, we define the perturbing potential to be \textit{periodic} if there is a positive integer $N$ and duration $T$ such that the time-dependent potential $\hat{V}(t)$ satisfies:
\begin{equation}
    \hat{V}(t + NT) = \hat{V}(t).
\end{equation}
Furthermore, we impose the \textit{cyclic} condition on $\hat{V}(t)$:
\begin{equation}
    \int_0^{NT} \hat{V}(t) dt = 0.
\end{equation}
$\hat{V}(t)$ may, in general, admit \textit{subcycles}, which means that for some $0 < k < N$: 
\begin{equation}
    \int_0^{kT} \hat{V}(t) dt = 0.
\end{equation}

\subsubsection{Operators as Linear Combinations of Pauli Matrices}
\label{sec:operatorslinearcomb}

We further assume that the underlying physical system is multipartite, and the indices appearing for each partition will be thought of as a site-indexing value. The perturbing potential can be generically represented as: 
\begin{equation}
    \hat{V}(t) = \sum_{\alpha \in \mathcal{A}} \sum_{j} h_{j}^{\alpha}(t) \hat{\sigma}_{j}^{(\alpha)},
    \label{eq:perturbingPotential}
\end{equation}
where $\mathcal{A} = \{1,2,3\}$ is an indexing set of the Pauli-matrices.  We also assume that the functions $h^{\alpha}_{j}(t)$ have the form:
\begin{equation}
    h^{\alpha}_{j}(t) = f^{\alpha}_{j}(t) + \omega g^{\alpha}_{j}(t),
    \label{eq:hIsfPlusg}
\end{equation}
thereby allowing us to separate $\hat{V}(t)$ into a sum of two terms: the term proportional to $f^{\alpha}_{j}(t)$ has an amplitude that is small compared to $\omega$ while the $\omega g^{\alpha}_{j}(t)$ term in general has an amplitude on the order of $\omega$. The case considered in Ref.~\onlinecite{goldman2014periodically} takes $g^{\alpha}_{j}(t) = 0$ for all $t$ (with some special considerations in the appendix); we generalize this assumption in this work (Appendix~\ref{app:GeneralizationProof} proves our generalization). Eq.~\eqref{eq:hIsfPlusg} allows us to access a large class of pulse profiles.

Figure~\ref{fig:PulseGeneralization} shows example pulse shapes encompassed by Eq.~\eqref{eq:perturbingPotential}.   Panel (a) displays conventional pulse profiles \cite{Vandersypen2005} where pulse shapes are assumed to be rectangular for mathematical convenience in performing expansion integrals.  The rectangular shape introduces a width time scale, $\tau_w$, implied by this type of pulse.  Other time scales include the inverse pulse strength and those arising from derivatives near the edge of the rectangle (not shown).  These time scales must be included in an accurate effective Hamiltonian formalism to mitigate errors.  Furthermore, attempts to approximate ideal square pulse profiles in the laboratory can also lead to errors.  Panel (b) shows another example class of pulse profiles where a discrete Fourier transform approximation to an otherwise continuous pulse profile is introduced.  Here, again, the timescale $\tau_w$ is introduced along with other approximations that must be accounted for.  Panel (c) depicts the general class of pulse profiles, Eq.~\eqref{eq:perturbingPotential}, allowed by our formalism that include panels (a) and (b) as subsets.  To accommodate pulses depicted in panel (c), our formalism must capture all assumptions regarding time scales and therefore allows pulses of arbitrary strength and duration provided $\omega^{-1}$ is kept as a small parameter.

\subsubsection{Form of the Kick Operator} 
\label{sec:timeordering}

We assume that the relationship between the kick operator and the perturbing potential is given by
\begin{equation}
    \hat{K}^{(0)}(t) = \int^t\hat{V}(t')dt' = \sum_{\alpha, j} G^{\alpha}_j(t) \hat{\sigma}^{(\alpha)}_j
    \label{eq:lowestKick}
\end{equation}
with
\begin{equation}
    G^{\alpha}_j(t) = \omega \int^tg^{\alpha}_j(t')dt'.
    \label{eq:Gfunction}
\end{equation}
One can show that this form of the kick operator can be obtained by requiring that the terms of order $\mathcal{O}(\omega)$ sum to zero in the full expansion for $\hat{H}_{\text{eff}}$. [See Appendix \ref{app:FullDerivation} and Eq.~\eqref{eq:FullEffectiveHamiltonian}.] 

\begin{figure}
    \includegraphics[width=1\linewidth]{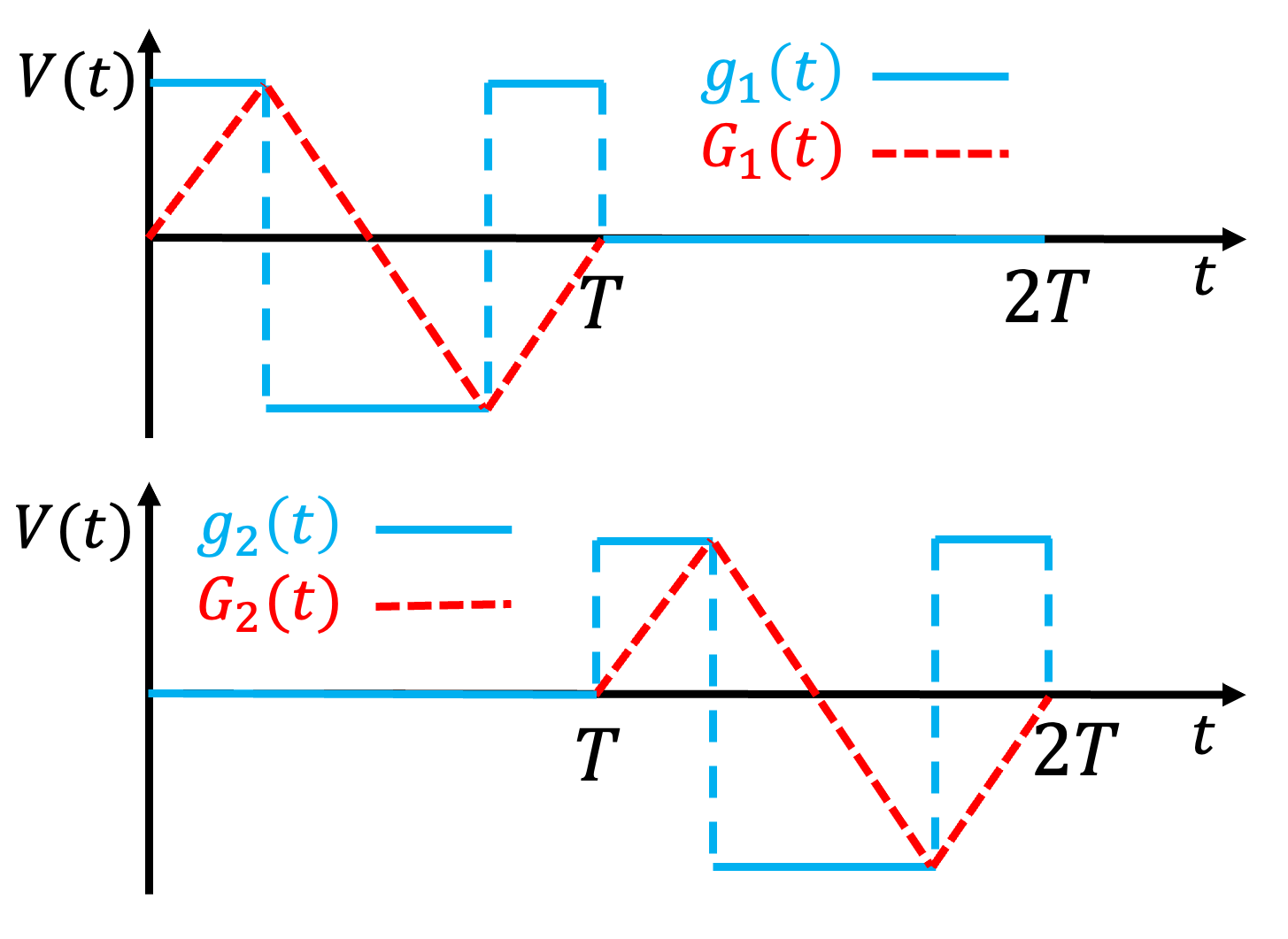}
    \caption{Example of a square wave pulse with cycle time $2T$.
    Top: The solid line shows $g_1(t)$ in the first subcycle: the global pulse profile for application of a $\sigma^{(1)}$ Pauli matrix. The dashed line shows the integral of the solid line, $G_1(t)$, defined by Eq.~\eqref{eq:Gfunction}.  Bottom: The same as the top but for a second subcycle pulsing the  $\sigma^{(2)}$ Pauli matrix.}
    \label{fig:squareWaveExample}
\end{figure}

We make simplifying assumptions for the pulse.  
As a special case, we assume that pulses are time ordered, expressed as a restriction on $g^{\alpha}_{j}(t)$. In general, this restriction is not necessary, but it simplifies our formalism.  For simplicity, we can consider the case where $f^{\alpha}_{j}(t) = 0$ for all indices. We model the time-ordering by imposing a mathematical condition on the coefficients which `turns on' one local spin operator at each site $j$ for each interval of $T$ so that we create subcycles of $[0,NT]$ of duration $T$. This can be expressed by requiring the $g^{\alpha}_{j}(t)$ to be piecewise continuous. Furthermore, we will also assume that the shape of the pulse remains the same as we cycle and also that the pulses are global. This means the functions have the form:
\begin{equation}
    g^{\alpha}_{j} (t)= \begin{cases}
        g(t) & t \in [(\alpha-1)T, \alpha T]\\
        0 & \text{Otherwise},
    \end{cases}
    \label{eq:PulseShape}
\end{equation}
where the pulse turns on and off the Pauli operators indexed by $\alpha$ in distinct time windows $t \in [(\alpha-1)T, \alpha T]$.  We choose the $g(t)$ to satisfy the conditions: i) $\int_{(\alpha-1)T}^{\alpha T} g(t)dt = 0$ which means that each $g^{\alpha}_{j}(t)$ has a subcycle time of $T$; ii) $G(t)$ has half-wave symmetry, so that $G(\frac{2\alpha+1}{2}T - \delta t) = -G(\frac{2\alpha+1}{2}T + \delta t)$, with $0 < \delta t < T/2$. An example time ordering is depicted in Fig.~\ref{fig:squareWaveExample} for $N = 2$ and $g(t)$ is a (translated) square wave.

Additionally, because we choose the same $g(t)$ for each of the $g^{\alpha}_{j}(t)$, the coefficients are not site-dependent and the form of the pulse as it cycles through the Pauli operators remains the same (although it is straightforward to relax this assumption in application of our formalism).  As a convention, whenever a function does not depend on site index, we use a subscript notation for indexing rather than superscript to avoid confusion with exponentials. For example, in  Eq.~\eqref{eq:PulseShape}, $g^{\alpha}_{j}(t)$ does not depend on site number, so we set $g^{\alpha}_{j}(t) \rightarrow g_{\alpha}(t)$. 

\subsubsection{Convergence Considerations} 
\label{sec:FrequecyRestrictions}

The general problem of determining the radius of convergence of high frequency expansions at infinite times is non-trivial and remains an open question (See Refs.~\onlinecite{rahav2003effective,goldman2014periodically,bukov2015universal,KUWAHARA2016} for discussions).   However, high frequency expansions can be safely applied in practice for sufficiently short total time scales $NT$ such that issues related to the radius of convergence \cite{MOAN2001a} are avoided.  In the following we therefore assume that, in practice, a quantum simulator evolves for a finite amount of time such that perturbative truncations discussed below are small and the system does not reach a long time asymptotic regime \cite{bukov2015universal,KUWAHARA2016}.

\subsection{Effective Hamiltonian from Strong Driving}
\label{sec:strongdriving}

This section uses the above assumptions to construct a formalism for pulse engineering effective Hamiltonians starting with Eq.~\eqref{eq:hamiltonianSystem}. We require the potential to be cyclic and periodic, as discussed in Sec.~\ref{sec:CyclicPeriodic}. The perturbing potential $\hat{V}(t)$ is assumed to have the form given in Sec.~\ref{sec:operatorslinearcomb}  and Sec.~\ref{sec:timeordering}, i.e., we assume Eqs.~\eqref{eq:perturbingPotential}-\eqref{eq:hIsfPlusg} and Eqs.~\eqref{eq:lowestKick}-\eqref{eq:PulseShape}.  We proceed via a gauge transformation on the Schr\"{o}dinger equation defined by Eqs.~\eqref{eq:gaugeTransformation}.  We expand $\hat{H}_{\text{eff}}$ and $\hat{K}(t)$ in powers of inverse frequency following our assumption in Sec.~\ref{sec:FrequecyRestrictions}:
\begin{align}
    \hat{H}_{\text{eff}} &= \sum_{n = 0}^{\infty} \omega^{-n} \hat{H}_{\text{eff}}^{(n)} \label{eq:hExpansion}\\
    \hat{K} &= \sum_{n = 0}^{\infty} \omega^{-n} \hat{K}^{(n)}.
    \label{eq:kExpansion}
\end{align}
If we employ our assumptions in Sec.~\ref{sec:Assumptions}, take the time-average of the effective Hamiltonian, and only expand to lowest order, $\omega^0$, we obtain terms proportional to the average value of $g^{\alpha}_{j}(t)G^{\beta}_{j'}(t)$ and the average value of $G^{\alpha}_{j}(t)G^{\beta}_{j'}(t)$. We can choose the $g^{\alpha}_{j}(t)$ so that:
%\begin{align}
%    g^{\beta}_{j'} \cdot G^{\alpha}_{j} &= 0 \nonumber \\
%    G^{\beta}_{j'} \cdot G^{\alpha}_{j} &= (G^{\alpha}_{j})^2 \delta_{jj'}\delta_{\alpha \beta}, 
%\end{align}
\begin{align}
    \overline{g^{\beta}_{j'} G^{\alpha}_{j}} &= 0 \nonumber \\
    \overline{ G^{\beta}_{j'} G^{\alpha}_{j}} &= (G^{\alpha}_{j})^2 \delta_{jj'}\delta_{\alpha \beta}, 
\end{align}
%where the notation $f \cdot g = \int_0^T f(t)g(t) dt$ denotes the usual inner product on the space of periodic functions 
where the overline denotes the time average value of a function: $\overline{f} = T^{-1}\int_0^T f(t) dt$,  
and we suppress the $t$-argument for notational convenience. For example, choosing $g_{j}^{\alpha}(t) = v\cos(\omega t)$ ($v$ a dimensionless parameter) when $t \in [(\alpha-1)T,\alpha T]$ and $g_{j}^{\alpha}(t) = 0$ otherwise satisfies this condition. 

\begin{figure*}
    \includegraphics[width=\linewidth]{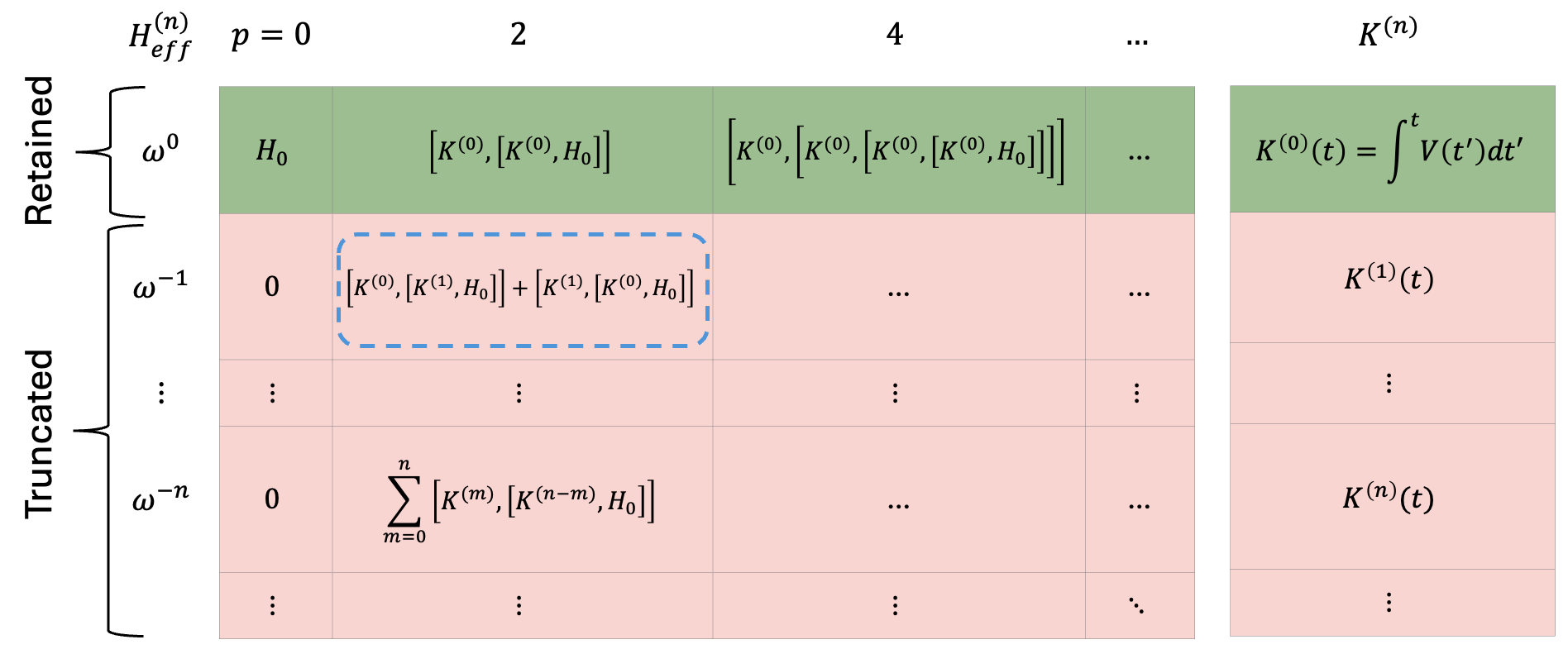}
    \caption{A schematic depicting the expansion procedure used to derive Eqs.~\eqref{eq:lowestKick} and ~\eqref{eq:effectiveEqn}.  The first row ($n=0$) in the central rectangle shows that Eq.~\eqref{eq:effectiveEqn} retains all nested commutators to lowest order in inverse frequency, $\omega^0$.  The rectangle on the right shows that the lowest order kick operator must also be kept.  The remaining rows ($n>0$) are truncated in our formalism.  The dashed line depicts an example $\mathcal{O}(\omega^{-1})$ correction to highlight the need for higher order expansions of the kick operator, $\hat{K}^{(1)}$, when retaining $n=1$ terms.}
    \label{fig:ExpansionTable}
\end{figure*}

Keeping all $\mathcal{O}(\omega^0)$ terms in the expansion yields the following time-averaged high frequency effective Hamiltonian (See Appendix~\ref{app:FullDerivation} for a derivation):
\begin{equation}
  \overline{\hat{H}_{\text{eff}}} = \hat{H}_0 + \sum_{j}\sum_{\alpha\in \mathcal{A}}F^{\alpha}_{j}[\hat{H}_0]+ \mathcal{O}(\omega^{-1}), 
    \label{eq:effectiveEqn}
\end{equation}
where we define the linear superoperator: 
\begin{equation}
    F^{\alpha}_{j}[\hat{H}] = \sum_{p=1}^{\infty} \frac{(-1)^p}{(2p)!}\overline{\left(G^{\alpha}_{j}\right)^{2p}} [[\hat{\sigma}^{(\alpha)}_{j},\hat{H}]]_{2p},
    \label{eq:Fseries}
    \end{equation}
where the notation $[[\hat{A},\hat{B}]]_{p}=[\hat{A},...[\hat{A},\hat{B}]...]$ indicates a nested commutator occurring $p$ times such that:
\begin{align}
[[\hat{A},\hat{B}]]_0&= \hat{B}, \nonumber\\ 
[[\hat{A},\hat{B}]]_1&=  [ \hat{A},\hat{B}], \nonumber\\ 
[[\hat{A},\hat{B}]]_2&=  [\hat{A}, [ \hat{A},\hat{B}]], \nonumber\\ 
[[\hat{A},\hat{B}]]_3&=  [\hat{A},[\hat{A}, [ \hat{A},\hat{B}]]], \nonumber\\ 
\vdots \nonumber
\end{align}
$F^{\alpha}_{j}[\hat{H}]$ enables us to think of the engineering process as a transformation on the operator algebra over the state space.  Eq.~\eqref{eq:effectiveEqn} must also be paired with the solution to the differential equation for $\hat{K}^{(0)}$, Eq.~\eqref{eq:lowestKick}, to compute observables.  Note that the effective Hamiltonian applies only in the time average sense and that this approach differs from the use of term-by-term cancellation of time dependence without averaging (see, e.g., Ref.~\cite{Bukov2016}). 

Equations~\eqref{eq:lowestKick} and ~\eqref{eq:effectiveEqn} establish central results. These equations, under the  assumptions discussed in Sec.~\ref{sec:Assumptions}, are exact to $\mathcal{O}(\omega^{0})$.  The green rectangular regions in Fig.~\ref{fig:ExpansionTable} show the terms retained in Eqs.~\eqref{eq:lowestKick} and ~\eqref{eq:effectiveEqn}.  Fig.~\ref{fig:ExpansionTable} also shows terms that need to be retained to include $\mathcal{O}(\omega^{-1})$ corrections in $\overline{\hat{H}_{\text{eff}}}$.  The dashed line highlights an example higher order correction to emphasize that rigorous inclusion of such terms requires solving for both $\hat{K}^{(0)}$ \textit{and} $\hat{K}^{(1)}$. Appendix~\ref{app:FullDerivation} also discusses the higher order $\mathcal{O}(\omega^{-1})$ corrections in the derivation.  Eqs.~\eqref{eq:lowestKick} and ~\eqref{eq:effectiveEqn} are also very general.  Specifically, Eqs.~\eqref{eq:lowestKick} and ~\eqref{eq:effectiveEqn} can be applied in a non-stroboscopic setting--without further assumptions regarding time scales--and to general spin models $H_0$ and, as a concrete example, models captured by Eq.~\eqref{eq:spinModels}.  

The infinite series in  Eq.~\eqref{eq:Fseries} must be evaluated to all orders to allow arbitrary pulse shapes.  But an alternative simple model arises by truncating the series in $p$.  Keeping only the $p=1$ term leads to:
\begin{equation}
    \overline{\hat{H}_{\text{eff}}} \approx \hat{H}_0 - \frac{1}{2}\sum_{j}\sum_{\alpha\in \mathcal{A}} \overline{(G^{\alpha}_{j})^2} \left[\hat{\sigma}^{(\alpha)}_{j},\left[\hat{\sigma}^{(\alpha)}_{j}, \hat{H}_0\right]\right],
    \label{eq:lowestorder}
\end{equation}
where the corrections in the expansion are proportional to $\overline{(G^{\alpha}_{j})^4}$.  As discussed above, pulse profile choices can implicitly introduce time scales and require retaining all $p$ but Eq.~\eqref{eq:lowestorder} will suffice if $\overline{(G^{\alpha}_{j})^4}$ and higher order corrections in $p$ converge rapidly.  But if they do not converge rapidly because non-trivial time scales are introduced by the pulse profiles (See, e.g., ~Ref.~\onlinecite{GEIER2021c}), there will be significant corrections to Eq.~\eqref{eq:lowestorder} that must be accounted for.  We now turn to the XXZ models where we sum the nested commutators in Eq.~\eqref{eq:Fseries} explicitly to derive analytic control conditions for $\overline{\hat{H}_{\text{eff}}}$ at all orders of $p$.

\section{Application to XXZ Spin Models}
\label{sec:xxmodel}

We now apply our formalism to XXZ spin models.   We then discuss example pulses and closed form solutions that allow effective Hamiltonian engineering of Eq.~\eqref{eq:spinModels}.  We focus on spatially global pulses as experimentally relevant examples. 

\subsection{General Conditions for Engineering}

We seek to engineer Eq.~\eqref{eq:spinModels} into different limits using Eqs.~\eqref{eq:lowestKick} and ~\eqref{eq:effectiveEqn}.  In this and the following section we take as the periodic perturbing potential:
\begin{equation}
    \hat{V}(t) = v\omega\left[g_1(t) \sum_{k} \hat{\sigma}^{(1)}_{k} + g_2(t) \sum_{k} \hat{\sigma}^{(2)}_{k}\right].
    \label{eq:VasSigmaXandSigmaY}
\end{equation}
 Here $v$ is a dimensionless parameter characterizing the ratio of the amplitude to frequency of the perturbing pulse.   Note that this pulse is linear in $\omega$ and is therefore a strong driving pulse.

 Appendix~\ref{app:fullDerivationIsing} details another central result of our work.  We insert Eq.~\eqref{eq:VasSigmaXandSigmaY} into Eq.~\eqref{eq:effectiveEqn} and evaluate all nested commutators to find solutions for the following effective models:
\begin{equation}
    \overline{\hat{H}^\text{S}_{\text{eff}}} = A\hat{H}_{XY} + B\hat{H}_{ZZ},
    \label{eq:EffHamIsAXYplusBZZ}
\end{equation}
where the (dimensionless) coefficients are: 
\begin{align}
\begin{split}
    A &= 1 - \frac{8 \delta J }{J_{\perp}}U  \\
    B &= 1 - \frac{16 \delta J }{J_z}U \\
    U &= \frac{1}{16} \sum_{p=1}^{\infty} \frac{(-1)^p (4v)^{2p}}{(2p)!} \overline{G^{2p}}. \label{eq:UFunction}
    \end{split}
\end{align}
Here $\delta J = J_z - J_{\perp}$ and $G = \omega\int^tg(t')dt'$.  $U$ arises as a consequence of the sum in Eq.~\eqref{eq:Fseries} and abstains from a particular choice for $g(t)$.  Implicitly, this means that the coefficients $A$ and $B$ depend on the form of the pulses chosen, which is to be expected (see Appendix~\ref{app:fullDerivationIsing}).  

Equation~\eqref{eq:UFunction}  can be used to engineer a model of the type: $A_c \hat{H}_{XY} + B_c\hat{H}_{ZZ}$, where $A_c $ and $B_c$ are user-defined constants.  
The problem reduces to choosing $g(t)$ to solve a system of equations given by: 
\begin{align}
    A = A_c \text{  and  } 
    B = B_c,
    \label{eq:ABconditions}
\end{align}
which specifies the effective Hamiltonian. Lastly, we remark that Eq.~\eqref{eq:EffHamIsAXYplusBZZ} is broad enough in $A$ and $B$ (by inspecting their forms) that one can engineer on $\hat{H}_{XY}$ or $\hat{H}_{XXZ}$ to produce a large class of effective Hamiltonians [see also Eq.~\eqref{eq:effH0XXZ}]. For example, for $A=0$, we obtain the special case: $\overline{\hat{H}^{\text{S}}_{\text{eff}}} \rightarrow \hat{H}_{ZZ}$.  For $B=0$, we obtain $\overline{\hat{H}^{\text{S}}_{\text{eff}}}\rightarrow \hat{H}_{XY}$ and for $A=B$, $\overline{\hat{H}^{\text{S}}_{\text{eff}}}$ becomes the Heisenberg model: $\sim \sum_{j \neq j'} \mathcal{V}_{j,j'}[\hat{\sigma}^{(1)}_{j}\hat{\sigma}^{(1)}_{j'} + \hat{\sigma}^{(2)}_{j} \hat{\sigma}^{(2)}_{j'} + \hat{\sigma}^{(3)}_{j}\hat{\sigma}^{(3)}_{j'}]$.

\subsection{Example Pulses}

We will consider example pulse structures and derive conditions that must be satisfied in order for the desired engineering outcome. In particular, we consider the case of a single global cosine wave and the case of a global square wave. In these cases we sum the nested commutators in Eq.~\eqref{eq:Fseries} analytically to obtain closed forms for $U$ but numerical truncations of Eq.~\eqref{eq:UFunction} are also possible.  We will also restrict to the case where we look to engineer an Ising interaction from an XXZ interaction.  

The condition required to transform Eq.~\eqref{eq:EffHamIsAXYplusBZZ} into an effective Ising interaction is expressed by the constraint $A = 0$. If we impose this condition and solve for $U$,  we find:
\begin{equation}
    U = \frac{1}{8(s-1)},
    \label{eq:IsingCondition}
\end{equation}
 where $s = J_z/J_{\perp}$.  Eq.~\eqref{eq:IsingCondition} relates the pulse shape $g(t)$ (particularly its strength, $v$) to the coefficients appearing in $\hat{H}_{XXZ}$. 

\subsubsection{Cosine Wave}

For the simplest case, let: 
\begin{equation}
    g_{\alpha}(t) = \begin{cases}
        \cos\left(\frac{2\pi}{T}t\right) & t \in [(\alpha-1)T, \alpha T]\\
        0 & \text{Otherwise}
        \label{eq:cosinepulse}
    \end{cases} 
\end{equation}
 which implies:
\begin{equation}
    G_{\alpha}(t) = \begin{cases}
        \sin\left(\frac{2\pi}{T}t\right) & t \in [(\alpha-1)T, \alpha T]\\
        0 & \text{Otherwise}.
    \end{cases}
    \label{eq:Fforcosine}
\end{equation}
This pulse produces a lowest order kick operator:
\begin{align}
        \hat{K}^{(0)}(t) &= v\sin\left(\frac{2\pi}{T}t\right)\Theta(t)\Theta(T-t)\sum_{k} \hat{\sigma}^{(1)}_{k} \nonumber \\
    &+ v\sin\left(\frac{2\pi}{T}t\right)\Theta(t-T)\Theta(2T-t) \sum_{k} \hat{\sigma}^{(2)}_{k},
\end{align}
where we use the Heaviside step function, $\Theta(t)$, to represent the piecewise continuous forms of $G_{\alpha}$. Using Eq.~\eqref{eq:Fforcosine} in Eq.~\eqref{eq:UFunction} leads to the closed form expression $U$ for the case of the cosine pulse:
\begin{align}
U_{\text{c}}= \frac{J_0(4v) - 1}{32},
\end{align}
where $J_0(z)$ is the $0^{th}$ Bessel function of the first kind.
Using Eq.~\eqref{eq:IsingCondition} we obtain a condition relating the pulse strength $v$ to $s$:
\begin{equation}
    J_0(4v) = \frac{s+3}{s-1}.
\end{equation}
 $J_0(z)$ takes on values (approximately) in the interval $[-0.4,1]$. If we determine the set of $s$ such that $(s+3)/(s-1)$ takes on values in this interval, we can determine the set of $\hat{H}_{XXZ}$ Hamiltonians that can be engineered into $\hat{H}_{ZZ}$ using Eq.~\eqref{eq:cosinepulse}. Doing this, we obtain the set $s \in (-\infty,-13/7]$. Note in particular as $s \rightarrow \pm\infty$, we have $v \rightarrow 0$, which makes sense since this corresponds to $J_{\perp} \rightarrow 0$, so that $\hat{H}_0 = \hat{H}_{ZZ}$. The restriction to $s<0$ implies that a cosine wave can produce an Ising model from the XXZ model \textit{only if there is a relative minus sign between $J_{\perp}$ and $J_z$}. 

\begin{figure}
    \centering
    \includegraphics[height=0.85\textheight, keepaspectratio]{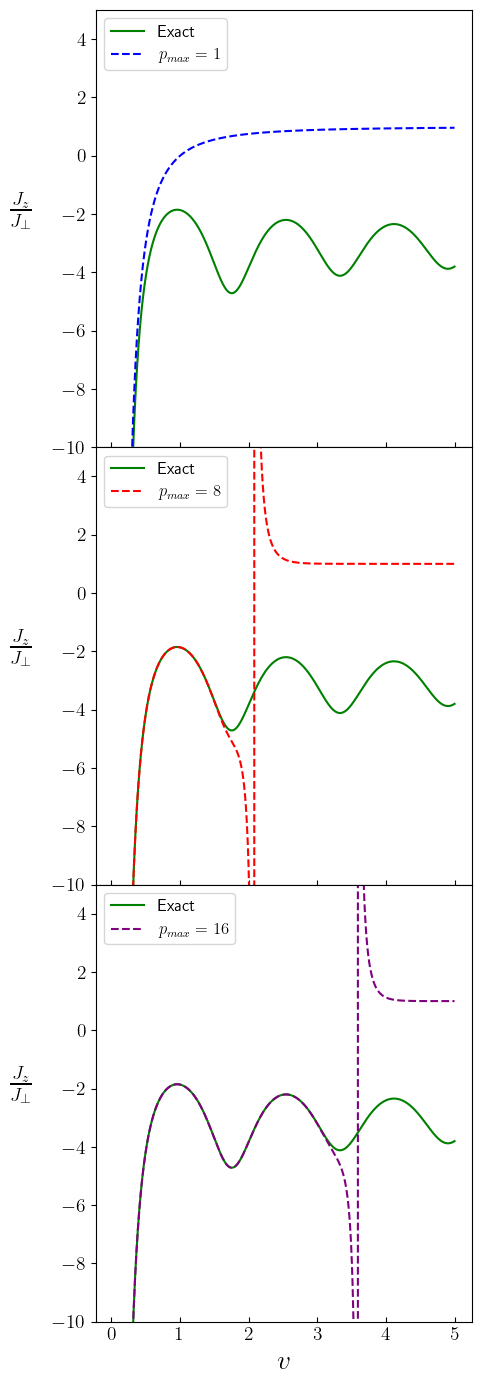}
    \caption{The solid lines plot the surfaces of solutions for $J_z/J_{\perp}$ obtained from Eq.~\eqref{eq:IsingCondition} plotted 
    against the pulse strength, $v$.  The pulse is assumed to be a cosine shape, Eq.~\eqref{eq:cosinepulse}. In the top, middle, and bottom panels the dashed lines plot the truncated approximations terminating at $p_{max} = 1$, $p_{max} = 8$, and $p_{max} = 16$, respectively. The vertical dashed lines indicate a pole in the truncated expansion.  As expected, when $v \rightarrow 0$, lower order approximations are good fits to the solution surface. On the other hand, every truncated solution incorrectly predicts the existence of values of $v$ for $s > 0$ (i.e., $J_z$ and $J_{\perp}$ having the same sign) which are pushed further and further out to $v \rightarrow \infty$. To determine all admissible values of $v$ which effectively engineer an Ising-type interaction from a given form of $H_{XXZ}$, we must keep \textit{all} terms in the expansion.}
    \label{fig:cosineSolutions}
\end{figure}

Appendix~\ref{app:physicalsystems} briefly reviews recent quantum simulation proposals and experiments realizing XXZ models with $\mathcal{V}_{j,j'}=\vert j-j'\vert^{-3}$.  Trapped circular Rydberg atoms can be engineered into an XXZ model where there can be a relative minus sign between $J_{\perp}$ and $J_z$ \cite{NGUYEN2018}. Polar molecules, however, defining pseudospins in the simplest setup (rotational states with total angular momenta $\mathcal{N}=0,1$ and zero $z$-component of angular momentum) do not always allow a relative minus sign between  $J_{\perp}$ and $J_z$.  We therefore find that Eq.~\eqref{eq:cosinepulse} does not allow a conversion from 
$H_{XXZ}$ to $H_{ZZ}$ for dipolar interacting molecules.  But this condition can be generically met in bipartite spin model approximations to trapped polar molecules, e.g., a nearest neighbor interaction \cite{Lee2016} $\mathcal{V}_{j,j'}\approx\delta_{j,j'\pm1}$ on a chain lattice, by performing a gauge transformation on every other site on the chain: $\sigma^{(1)}_{2j}\rightarrow -\sigma^{(1)}_{2j}$ and $\sigma^{(2)}_{2j}\rightarrow -\sigma^{(2)}_{2j}$.  The gauge transformation allows a relative minus sign between $J_{\perp}$ and $J_{z}$ (on a bipartite model).  We therefore conclude that for the global cosine pulse, dipolar XXZ models engineered by circular Rydberg atom arrays can be driven into an Ising regime whereas only bipartite models of interacting $\mathcal{N}=0,1$ Alkali-based polar molecules can be converted from an XXZ model to an Ising model. 

Figure~\ref{fig:cosineSolutions} uses a strong driving cosine pulse to highlight the need to retain all orders in $p$ in Eq.~\eqref{eq:Fseries} in a broad parameter search designed to engineer models. One could in principle use the $p_{max} = 1$ term, i.e., Eq.~\eqref{eq:lowestorder}, to attempt to engineer the XXZ model into an Ising model, but this leads to qualitatively incorrect results in a pulse parameter sweep. In particular, the first-order solution to Eq.~\eqref{eq:IsingCondition} suggests a smooth set of values for $v$ to select from that effectively engineers all possible ratios of $J_z$ to $J_{\perp}$ up to $J_z = J_{\perp}$, as seen by the top panel in Fig.~\ref{fig:cosineSolutions}. However, even truncating at large but \textit{finite} $p_{max}$ still predicts the existence of values of $v$ which could in principle engineer specific forms of the $H_{XXZ}$, but as one keeps more and more terms in  Eq.~\eqref{eq:UFunction}, these solutions are pushed out to $v \rightarrow \infty$. For large $v$, the solutions are \textit{non-physical} and this can only be demonstrated by keeping all orders in $p$ in wide parameter searches over all $v$.   

We argue qualitatively for the need to retain all $p$ as follows: for an arbitrary pulse shape, there is no obvious small parameter in each term in Eq.~\eqref{eq:Fseries}; attempting to use a finite set of terms as in Eq.~\eqref{eq:lowestorder}, one predicts a $v$ which may seem plausible but is not obviously `small'. Consequently, for the derived $v$ the $p>1$ terms in Eq.~\eqref{eq:Fseries} oscillate in sign and converge only for large $p$. As these higher order terms are kept, the supposed solution becomes contradictory as seen by our calculations above.  Fig.~\ref{fig:cosineSolutions} shows that if we employ Eq.~\eqref{eq:lowestorder} (i.e., $p_{max} = 1$) to predict a value for $v$ assuming $s > 0$, we find $v \gtrsim 1$. But then if we look at the cases for $p_{max} = 8$ and $p_{max} = 16$, we can see that the selected solution \textit{ceases to exist} as we keep more and more terms in Eq.~\eqref{eq:UFunction}. In wide parameter sweeps of $v$ we must therefore include all possible terms in Eq.~\eqref{eq:UFunction}.

\subsubsection{Square Wave}

We also consider a conventional square wave pulse of the form:
\begin{align}
    g_{\alpha}(t) = \begin{cases}
        1 & t \in \left[(\alpha-1)T, \left(\frac{4\alpha-3}{4}\right)T\right] \cup \left[\left(\frac{4\alpha - 1}{4}\right)T,\alpha T\right]  \\
        -1 & t \in \left[\left(\frac{4\alpha-3}{4}\right)T,\left(\frac{4\alpha - 1}{4}\right)T\right] \\
        0 & \text{Otherwise}, \nonumber 
    \end{cases}
\end{align}
as depicted in Fig. \ref{fig:squareWaveExample}. This choice implies that the $G_{\alpha}$ are triangle waves. Using this new pulse form the infinite sum in Eq.~\eqref{eq:UFunction} gives:
\begin{equation}
    U_{\text{s}} = \frac{1}{16}\left[\frac{\sin(2 \pi v)}{2 \pi v} - 1\right].
\end{equation}
Substituting into Eq.~\eqref{eq:IsingCondition}  gives the condition for engineering an Ising interaction with the square pulse:
\begin{equation}
    \frac{\sin(2 \pi v)}{2 \pi v} = \frac{s + 1}{s -1}.
\end{equation}
Analyzing in the same way as the cosine pulse case, we can find the domain of $s$ such that this equation is satisfied. We first note again that as $s \rightarrow \pm \infty$, the only solution is $v = 0$, which makes sense. On the other hand, the finite value for which there is a solution is when the left-hand side has a minimum since there are no finite solutions for  $s > 1$ following from the fact that $\sin(z)/z$ is bounded above by 1. This occurs first at $v \approx -0.715$ with a value of $\sin(2\pi v)/(2\pi v) \approx -0.217$. This yields $s \in (-\infty, -0.643]$. Again for a square wave, $s < 0$ so that the relative sign between $J_{\perp}$ and $J_z$ must be negative.

\section{Summary and Outlook}
\label{sec:summary}

We have constructed a formalism that loosens the set of assumptions \cite{goldman2014periodically} underlying pulse shapes for Floquet engineering of interacting spin systems into accurate effective Hamiltonians.   Our formalism allows tunability of pulse shape, e.g., width and strength, while maintaining controlled error.  Specific examples using global Pauli-$x$ and Pauli-$z$ pulses were implemented to demonstrate conversion of long-range XXZ spin models to Ising models.  The formalism constructed here is accurate to lowest order in inverse frequency of the driving pulses, $\mathcal{O}(\omega^0)$.  It would be interesting to compare our Floquet engineering formalism to digital Trotter-based methods (See, e.g., Refs.~\cite{PASTORI2022,CHERTKOV2024,ECKSTEIN2024,HAGHSHENAS2025}).

The method discussed here can be further generalized.  i) We focused on pulse profiles with summation formulas for $F^{\alpha}_{j}[\hat{H}]$ in Eq.~\eqref{eq:Fseries} that yielded analytically tractable conditions for Hamiltonian engineering.  It is straightforward to use our results for other pulse profiles that can be numerically summed to very large $p$ without the need for analytic summation.  ii) We examined global pulses but our formalism can be used to study local (site-dependent) pulses that allow wider tunability of effective Hamiltonians.  iii) Expansion to higher order, e.g., $\mathcal{O}(\omega^{-1})$, will also allow additional freedom in engineering effective models.  

It is also straightforward to adapt the formalism discussed here to models containing any operators respecting the Lie algebra commutation relations, not just $SU(2)$ spin systems.  Future work will allow Floquet-engineering of more general multi-component $SO(N)$ models relevant to rotational states of polar molecules (See Appendix~\ref{app:physicalsystems}) or $SU(N)$ models relevant to other states of molecules \cite{HOMEIER2024,MUKHERJEE2025} or superconducting qudits \cite{Kiktenko2025}.   In such cases, $F^{\alpha}_{j}[\hat{H}]$ will be modified to include new structure constants in commutation relations.   Interesting models that could be pursued include the Affleck–Kennedy–Lieb–Tasaki \cite{Affleck1987a} models which have interesting valence-bond solid ground states. 

\begin{acknowledgments}
We acknowledge support from AFOSR (FA9550-19-1-0272).  BG was supported by the AFOSR MURI program under agreement number FA9550-221-0339.  RS and VS acknowledge support from AFOSR (FA2386-21-1-4081, and FA9550-23-1-0034) and ARO (W911NF2210247). We thank B. DeMarco and S. Kotochigova for valuable conversations. 
\end{acknowledgments}

\appendix

\section{Proof of Generalization} \label{app:GeneralizationProof}

In this section we prove that we generalize the family of pulses defined by Eq.~K1 in Ref.~\onlinecite{goldman2014periodically}:
\begin{equation}
    \hat{V}^{\text{GD}}(t) = f(t)\hat{\mathcal{C}} + \omega g(t) \hat{\mathcal{B}},
    \label{eq:goldmanpotential}
\end{equation}
where $\hat{\mathcal{C}}$ and $\hat{\mathcal{B}}$ are combinations of Pauli operators.  We introduced Eq.~\eqref{eq:perturbingPotential} to generalize Eq.~\eqref{eq:goldmanpotential}.  In Eq.~\eqref{eq:perturbingPotential}, each local observable can be pulsed using an arbitrary pulse profile without introducing additional timescales in, e.g., a Fourier decomposition of the sought after pulses with a train of pulses much shorter than $T$.  Eq.~\eqref{eq:goldmanpotential}, by contrast, assumes that the pulse applies the same operator for a given pulse profile.  To distinguish between Eq.~\eqref{eq:goldmanpotential} and Eq.~\eqref{eq:perturbingPotential} we must show that our generalization contains all the pulses considered in Eq.~\eqref{eq:goldmanpotential} while also giving an example that cannot be reconstructed using Eq.~\eqref{eq:goldmanpotential} without introducing additional time scales. 

We first prove that Eq.~\eqref{eq:perturbingPotential} contains Eq.~\eqref{eq:goldmanpotential}. To do this, without loss of generality, we subsume the site indices $j$ in Eq.~\eqref{eq:perturbingPotential} into the index $\alpha$, following from the fact that a finite tensor product of finite dimensional Lie algebras is again a finite dimensional Lie algebra. Letting $\mathcal{A}$ be a finite set indexing all the generators, we have: 
\begin{equation}
    \hat{V}(t) = \sum_{\alpha \in \mathcal{A}} h_{\alpha}(t) \hat{\sigma}^{(\alpha)}.
    \label{eq:Vforproof}
\end{equation}
Since the Lie algebra is spanned by the $\{\hat{\sigma}^{(\alpha)}\}$, we have: 
\begin{align}
    \hat{\mathcal{C}} &= \sum_{\alpha \in \mathcal{A}} c_{\alpha}\hat{\sigma}^{(\alpha)}\\
    \hat{\mathcal{B}} &= \sum_{\alpha \in \mathcal{A}} b_{\alpha}\hat{\sigma}^{(\alpha)}
\end{align}
where the $b_{\alpha}$ and $c_{\alpha}$ are constants. We retrieve Eq.~\eqref{eq:goldmanpotential} by putting $h_{\alpha}(t) = f(t) b_{\alpha} + \omega g(t) c_{\alpha}$.

We now show that Eq.~\eqref{eq:Vforproof} allows example pulses not captured by Eq.~\eqref{eq:goldmanpotential}.  Consider an example pulse:
\begin{equation}
    \hat{V}(t) = \omega[\tilde{g}_1(t) \hat{\sigma}^{(1)} + \tilde{g}_2(t) \hat{\sigma}^{(2)}],
    \label{eq:pulsecounterexample}
\end{equation}
with:
\begin{align}
    \tilde{g}_1(t) &= \begin{cases}
        1 & 0 \leq t < T/4\\
        -1 & T/4 \leq t < 3T/4\\
        1 & 3T/4 \leq t < T\\
        0 & T \leq t < 2\pi \nonumber \\
    \end{cases}\\
    \tilde{g}_2(t) &= \tilde{g}_1(t-T) \nonumber.
\end{align}
In this case, we would have $f(t) = 0$ [a special case of Eq.~\eqref{eq:goldmanpotential}]. Additionally, for Eq.~\eqref{eq:goldmanpotential} to describe a pulse of this form, we would need: 
\begin{equation}
    \omega g(t) \sum_{\alpha \in \mathcal{A}}b_{\alpha}\hat{\sigma}^{(\alpha)} = \omega \sum_{\alpha \in \mathcal{A}} g_{\alpha}(t) \hat{\sigma}^{(\alpha)}.
\end{equation}
Since the $\{\hat{\sigma}^{(\alpha)}\}$ span the Lie algebra, this implies:
\begin{equation}
    \sum_{\alpha \in \mathcal{A}}[g(t)b_{\alpha} - g_{\alpha}(t)]\hat{\sigma}^{(\alpha)} = 0,
\end{equation}
and since the elements of the basis of the Lie algebra are all linearly independent,
$
    g(t)b_{\alpha} - g_{\alpha}(t) = 0, \hspace{0.2cm}$ for all $\alpha$.
But then this requires that: 
\begin{equation}
    \tilde{g}_1(t)/b_1 = g(t) = \tilde{g}_2(t)/b_2,
\end{equation}
or, equivalently:
\begin{equation}
    \frac{\tilde{g}_1(t)}{\tilde{g}_2(t)} = const.,
\end{equation}
which is a contradiction. We conclude that the pulse defined by Eq.~\eqref{eq:pulsecounterexample} is not of the form $\omega g(t) \hat{\mathcal{B}}$, and we have therefore shown that our generalization, Eq.~\eqref{eq:perturbingPotential}, encompasses a larger class of pulses than Eq.~\eqref{eq:goldmanpotential}.

\section{Full Derivation of the Lowest Order Hamiltonian and Corrections} 
\label{app:FullDerivation}

In this section we derive Eq.~\eqref{eq:effectiveEqn}. 
As mentioned in Sec.~\ref{sec:Formalism}, we consider a Hamiltonian and perturbing potential defined by  Eq.~\eqref{eq:perturbingPotential} where $\hat{V}(t)$ will be periodic and cyclic in the sense that for all the $g^{\alpha}_{j}(t)$:
\begin{align}
    g^{\alpha}_{j}(t + NT) = g^{\alpha}_{j}(t)\\
    \int_0^{NT}g^{\alpha}_{j}(t) dt = 0,
\end{align}
where $NT$ is the cycle time. Now consider the gauge transformation in Eq.~\eqref{eq:gaugeTransformation}, which gives rise to the effective Schr\"{o}dinger equation:  $\hat{H}_{\text{eff}} \phi = i \partial_t \phi$.
Furthermore, if the effective timescale of interaction of the Hamiltonian $\hat{H}_0$ is much larger than the timescale set by $\omega^{-1}$, we can use Eqs.~\eqref{eq:hExpansion} and \eqref{eq:kExpansion}.  We substitute Eqs.~\eqref{eq:hExpansion} and ~\eqref{eq:kExpansion} into Eqs.~\eqref{eq:gaugeTransformation} and use the identities:
\begin{align}
    e^{i\hat{K}(t)} \hat{H} e^{-i\hat{K}(t)} &= \sum_{p=0}^{\infty} \frac{i^p}{p!} [[\hat{K},\hat{H}]]_p\\
    \left(\frac{\partial e^{i\hat{K}(t)}}{\partial t}\right)e^{-i \hat{K}(t)} &= \sum_{p=0}^{\infty} \frac{i^{p+1}}{(p+1)!} \left[\left[\hat{K},\frac{\partial \hat{K}}{\partial t}\right]\right]_{p},
\end{align}
%where $[[\hat{A},\hat{B}]]_p = [\hat{A},...[\hat{A},\hat{B}]...]$ is a nested commutator occurring $p$ times in each of the sums such that
The substitution gives rise to the full expansion:

\begin{widetext}
    \begin{align}
        \sum_{n=0}^{\infty} \omega^{-n} \hat{H}_{\text{eff}}^{(n)} &= \sum_{p=0}^{\infty} \frac{i^p}{p!} \sum_{n_1,...,n_p = 0}^{\infty} \omega^{-(n_1 + ... + n_p)}[\hat{K}^{(n_1)},...[\hat{K}^{(n_p)},\hat{H}_0]...] \nonumber \\
        &+ \omega \sum_{\alpha} \sum_{j} g^{\alpha}_{j}(t) \left\{\sum_{p=0}^{\infty} \frac{i^p}{p!} \sum_{n_1,...,n_p = 0}^{\infty} \omega^{-(n_1 + ... + n_p)}[\hat{K}^{(n_1)},...[\hat{K}^{(n_p)},\hat{\sigma}^{(\alpha)}_{j}]...] \right\} \nonumber \\
        &+ i\sum_{q=0}^{\infty}\sum_{p=0}^{\infty} \omega^{-q}\frac{i^{p+1}}{(p+1)!}\left\{ \sum_{n_2,...,n_p=0}^{\infty} \omega^{-(n_2 + ... + n_p)}\left[\hat{K}^{(n_2)},...\left[\hat{K}^{(n_p)},\frac{\partial \hat{K}^{(q)}}{\partial t}\right]...\right]\right\}.
        \label{eq:FullEffectiveHamiltonian}
    \end{align}
\end{widetext}
In each sum, the commutator occurs $p$ times. Assume the lowest order term in the expansion of the kick operator satisfies:
\begin{equation}
    \hat{K}^{(0)}(t) = \sum_{\alpha,j}G^{\alpha}_{j}(t) \hat{\sigma}^{(\alpha)}_{j},
    \label{eq:AppBK0Soln}
\end{equation}
where:
\begin{equation}
    G^{\alpha}_{j}(t) = \omega \int^t dt' g^{\alpha}_j(t').
\end{equation}
This follows by requiring that the terms of order $\mathcal{O}(\omega)$ sum to zero on the right-hand side of the full expansion for $H_{\text{eff}}$ (Compare Appendix K in Ref.~\cite{goldman2014periodically}).

Returning to Eq.~\eqref{eq:FullEffectiveHamiltonian}, the lowest order of the effective Hamiltonian becomes:
\begin{equation}
    \hat{H}^{(0)}_{\text{eff}} = \hat{H}_0 + \sum_{p=1}^{\infty} \frac{i^p}{p!}[[\hat{K}^{(0)},\hat{H}_0]]_p + \hat{R},
    \label{eq:AppLowestOrder}
\end{equation}
where $\hat{R}$ is a remainder term that depends on both $\hat{K}^{(0)}$ and $\hat{K}^{(1)}$. In particular,
\begin{widetext}
    \begin{align}
        \hat{R} &= \sum_{p = 1}^{\infty} \frac{i^p}{p!}\sum_{n_1,...,n_p = 0}^{1}\delta_{\sum_{j=1}^p n_j,1}\left[\hat{K}^{(n_1)},...\left[\hat{K}^{(n_p)},\sum_{\alpha,j} g^{\alpha}_{j}(t) \hat{\sigma}^{(\alpha)}_{j}\right]....\right] 
        \nonumber 
        \\
        & + \frac{i}{\omega}\sum_{p=1}^{\infty} \frac{i^{p+1}}{(p+1)!} \sum_{n_1,...,n_p = 0}^{1}\delta_{\sum_{j=1}^p n_j,1}\left[\hat{K}^{(n_1)},...\left[\hat{K}^{(n_p)},\frac{\partial \hat{K}^{(0)}}{\partial t}\right]...\right] + \frac{i}{\omega}\sum_{p=0}^{\infty}\frac{i^{p+1}}{(p+1)!}\left[\left[\hat{K}^{(0)},\frac{\partial \hat{K}^{(1)}}{\partial t}\right]\right]_p 
        \nonumber \\
        &=   \sum_{p = 1}^{\infty} \frac{i^p}{p!} \frac{p}{p+1} \sum_{n_1,...,n_p = 0}^{1}\delta_{\sum_{j=1}^p n_j,1}\left[\hat{K}^{(n_1)},...\left[\hat{K}^{(n_p)},\sum_{\alpha,j} g^{\alpha}_{j}(t) \hat{\sigma}^{(\alpha)}_{j}\right]....\right]
        + \frac{i}{\omega}\sum_{p=0}^{\infty}\frac{i^{p+1}}{(p+1)!}\left[\left[\hat{K}^{(0)},\frac{\partial \hat{K}^{(1)}}{\partial t}\right]\right]_p \label{eq:remainder}
    \end{align}
\end{widetext}
where $\delta_{\sum_jn_j,1}$ means when the sum of the set $\{n_j\}$ is one, the delta function is one, and is zero otherwise.  The commutator occurs $p$ times in each term.  

To prove Eq.~\eqref{eq:effectiveEqn}, the remainder term (or its time average) must vanish.  Importantly, the strong driving forces the remainder term to depend on $\hat{K}^{(1)}$ so that, in general, the lowest order term in the effective Hamiltonian also appears to depend on $\hat{K}^{(1)}$.  If we can find a $\hat{K}^{(1)}$ such that  $\hat{R} = 0$,  $\hat{K}^{(1)}$ will not play a role in $\overline{\hat{H}^{(0)}_{\text{eff}}}$ but will instead appear as a gauge choice in Eq.~\eqref{eq:gaugeTransformation}.  We will then recover the results presented in Sec.~\ref{sec:xxmodel}.  A different condition, using the average of the remainder, could also be imposed: $\overline{\hat{R}} = 0$.  In the following, we choose the condition $\hat{R} = 0$ instead of  $\overline{\hat{R}} = 0$ to avoid additional assumptions on pulse profiles. 

We show two different solution methods for obtaining a vanishing remainder. The first is a trivial solution, $\hat{K}^{(1)}=0$ obtained by direct inspection of 
Eq.~\eqref{eq:remainder} where we see that all  terms contain one $\hat{K}^{(1)}$ or its derivative.  This proves that, for the simplest gauge choice, $\hat{K}^{(1)}=0$, Eq.~\eqref{eq:effectiveEqn} is correct.  

We now establish a second, more general protocol to generate solutions to $\hat{R}=0$.  We start by recalling that all observables are spanned by tensor products of Pauli matrices and the identity.  $\hat{K}^{(1)}$ must then admit the form:
\begin{equation}
    \hat{K}^{(1)} = \sum_{\vec{\alpha}} c_{\vec{\alpha}}(t)\hat{\sigma}^{(\alpha_1)}_1 \cdots \hat{\sigma}^{(\alpha_S)}_S = \sum_{\vec{\alpha}}c_{\vec{\alpha}}(t)\prod_j\hat{\sigma}^{(\alpha_j)}_j
    \label{eq:k1Form}
\end{equation}
where $\vec{\alpha}=\alpha_1,...,\alpha_S$ and $S$ is the number of spins in the system. If the $c_{\vec{\alpha}}(t)$ are determined, then so is $\hat{K}^{(1)}$. We now substitute Eqs.~\eqref{eq:AppBK0Soln} and ~\eqref{eq:k1Form} into Eq.~\eqref{eq:remainder} and set $\hat{R} = 0$. To do this, we will need to sum over a permutation of the locations of the product of Pauli operators appearing in \eqref{eq:k1Form}. So let $\mathcal{P}$ denote a permutation of the sequence of $S-1$ zeros and $1$ one, then we rewrite terms in Eq.~\eqref{eq:remainder}.  For the first term in Eq.~\eqref{eq:remainder} we can use the following:
\begin{widetext}
    \begin{align}
        &\sum_{\alpha,j} g^{\alpha}_{j}\delta_{\sum_{j=1}^p n_j,1}\left[\hat{K}^{(n_1)},...\left[\hat{K}^{(n_p)}, \hat{\sigma}^{(\alpha)}_{j}\right]....\right] \\ \nonumber
        &= \sum_{\alpha,j}\sum_{\substack{\beta_1,...,\beta_{p-1} \\ k_1,...,k_{p-1}}} \sum_{\vec{\gamma}}g^{\alpha}_{j}G^{\beta_1}_{k_1}\cdots G^{\beta_{p-1}}_{k_{p-1}}c_{\vec{\gamma}}\sum_{\mathcal{P}}\left[_{\mathcal{P}(0)}\hat{\sigma}^{(\beta_1)}_{k_1},...\left[_{\mathcal{P}(1)}\prod_{l}\hat{\sigma}^{(\gamma_l)}_l,...\left[_{\mathcal{P}(0)}\hat{\sigma}^{(\beta_{p-1})}_{k_{p-1}},\hat{\sigma}^{(\alpha)}_j\right]\right]\right].
    \end{align}
\end{widetext}
Whenever we consider a non-trivial permutation, the operator appearing just to the right of it is also permuted. With this substitution, Eq.~\eqref{eq:remainder} becomes:
\begin{widetext}
    \begin{align}
        \hat{R} &= \sum_{p=1}^{\infty}\frac{i^p}{p!}\frac{p}{p+1}\sum_{\alpha,j}\sum_{\substack{\beta_1,...,\beta_{p-1} \\ k_1,...,k_{p-1}}} \sum_{\vec{\gamma}}g^{\alpha}_{j}G^{\beta_1}_{k_1}\cdots G^{\beta_{p-1}}_{k_{p-1}}c_{\vec{\gamma}}\sum_{\mathcal{P}}\left[_{\mathcal{P}(0)}\hat{\sigma}^{(\beta_1)}_{k_1},...\left[_{\mathcal{P}(1)}\prod_{l}\hat{\sigma}^{(\gamma_l)}_l,...\left[_{\mathcal{P}(0)}\hat{\sigma}^{(\beta_{p-1})}_{k_{p-1}},\hat{\sigma}^{(\alpha)}_j\right]\right]\right] \label{eq:fullSigmaRemainder}\\
        &- \frac{1}{\omega}\sum_{p=1}^{\infty}\frac{i^p}{(p+1)!}\sum_{\substack{\beta_1,...,\beta_{p} \\ k_1,...,k_{p}}} \sum_{\vec{\gamma}}G^{\beta_1}_{k_1}\cdots G^{\beta_{p}}_{k_{p}}\frac{dc_{\vec{\gamma}}}{dt}\left[\hat{\sigma}^{(\beta_1)}_{k_1},...\left[\hat{\sigma}^{(\beta_{p})}_{k_{p}},\prod_{l}\hat{\sigma}^{(\gamma_l)}_l\right]\right],  \nonumber 
    \end{align}
\end{widetext}
where we have dropped the time dependence on $c_{\vec{\gamma}}$ for notational convenience.  The expression contains long Pauli strings.  To arrange them, we consider a permutation $\mathcal{P}_0$, so that the product $\prod_l \hat{\sigma}^{(\gamma_l)}_l$ appears at the $s^{th}$ position in the nested commutator, and $1 \leq s \leq p$.  One can also show that:
\begin{widetext}
    \begin{align}
        &\left[\hat{\sigma}^{(\beta_1)}_{k_1},...\left[\hat{\sigma}^{(\beta_{s-1})}_{k_{s-1}},\left[\prod_l\hat{\sigma}^{(\gamma_l)}_l,\left[\hat{\sigma}^{(\beta_{s+1})}_{k_{s+1}},...\left[\hat{\sigma}^{(\beta_p)}_{k_p},\hat{\sigma}^{(\alpha)}_j\right]\right]\right]\right]\right] \nonumber \\ 
        &= \sum_{\lambda}\left((-1)^{s-1}(2i)^p\sum_{\delta_1,...,\delta_{p-1}}\left(\prod_{r=1}^{p-1}\varepsilon^{\beta_{p+1-r}\delta_{r-1}\delta_r}\right)\varepsilon^{\beta_1\delta_{p-1}\lambda}\right)\left(\prod_{l \neq j}\hat{\sigma}^{(\gamma_l)}_l\right)\hat{\sigma}^{(\lambda)}_j, \label{eq:nestedCommutator}
    \end{align}
\end{widetext}
where we define $\delta_0 = \alpha$ and $\beta_s = \gamma_j$ (which is valid since $1 \leq s \leq p$ and the index on $\beta$ includes this range).  Here $\varepsilon^{\alpha \beta \gamma}$ denotes the usual Levi-Civita symbol.   We then use \eqref{eq:nestedCommutator} to simplify \eqref{eq:fullSigmaRemainder}. We exchange the sums over $\alpha, j, \vec{\gamma}, \lambda$ with the sum over $p$ since they are independent, to obtain:
\begin{equation}
    \hat{R} = \sum_{\substack{\alpha, j \\ \vec{\gamma}, \lambda}}\left(A^{\alpha\lambda\vec{\gamma}} g^{\alpha}_jc_{\vec{\gamma}} -\frac{B^{\alpha\lambda\vec{\gamma}}}{\omega}\frac{dc_{\vec{\gamma}}}{dt} \right)\left(\prod_{l \neq j}\hat{\sigma}^{(\gamma_l)}_l\right)\hat{\sigma}^{(\lambda)}_j,
    \label{eq:completeFormR}
\end{equation}
where the coefficients are
\begin{widetext}
    \begin{align}
        A^{\alpha\lambda\vec{\gamma}} &\equiv \sum_{p=1}^{\infty}\frac{i^{2p}}{p!}\frac{2^p p}{p+1}\sum_{\substack{\beta_1,...,\beta_{p-1} \\ k_1,...,k_{p-1}}}G^{\beta_1}_{k_1}\cdots G^{\beta_{p-1}}_{k_{p-1}}\sum_{s=1}^{p}\left((-1)^{s-1}\sum_{\delta_1,...,\delta_{p-1}}\left(\prod_{\substack{r=2 \\ r \neq p+1 - s}}^{p-1}\varepsilon^{\beta_{p+1-r}\delta_{r-1}\delta_r}\right)\varepsilon^{\beta_p\alpha\delta_1}\varepsilon^{\gamma_j\delta_{p-s}\delta_{p+1-s}}\varepsilon^{\beta_1\delta_{p-1}\lambda}\right),\\
        B^{\alpha\lambda\vec{\gamma}} &\equiv \sum_{p=1}^{\infty}\frac{i^{2p}(-2)^p}{(p+1)!}\sum_{\substack{\beta_1,...,\beta_p \\ k_1,...,k_p}}G^{\beta_1}_{k_1}\cdots G^{\beta_{p}}_{k_{p}}\sum_{\delta_1,...,\delta_{p-1}}\left(\prod_{r=2}^{p-1}\varepsilon^{\beta_{p+1-r}\delta_{r-1}\delta_r}\right)\varepsilon^{\gamma_j\alpha\delta_{1}}\varepsilon^{\beta_1\delta_{p-1}\lambda}.
    \end{align}
\end{widetext}
Since the generators are linearly independent, the condition $\hat{R} = 0$ with the form in Eq. \eqref{eq:completeFormR} gives as many equations as there are undetermined coefficients. These produce a set of coupled differential equations that can be solved generally to give a class of solutions for $\hat{K}^{(1)}$. That is, we must solve:
\begin{equation}
    A^{\alpha\lambda\vec{\gamma}} g^{\alpha}_jc_{\vec{\gamma}} -\frac{B^{\alpha\lambda\vec{\gamma}}}{\omega}\frac{dc_{\vec{\gamma}}}{dt} = 0, \hspace{0.25cm} \text{all $\vec{\gamma}$}.
    \label{eq:K1Coefficients}
\end{equation}
There are solutions to Eqs.~\eqref{eq:K1Coefficients}.  The simplest solution is $c_{\vec{\gamma}} = 0$, thus confirming that $\hat{K}^{(1)}=0$ is a valid gauge choice as discussed above.  Different choices for $\hat{K}^{(1)}$ will, in general, require the construction of formal solutions to the set \eqref{eq:K1Coefficients}.  In each case, the vanishing remainder term leads to:
\begin{equation}
    \overline{\hat{H}}_{\text{eff}} = \hat{H}_0 + \sum_{p=1}^{\infty} \frac{i^p}{p!}\overline{[[\hat{K}^{(0)},\hat{H}_0]]_p} + \mathcal{O}(\omega^{-1}), 
\end{equation}
which leads to Eq.~\eqref{eq:effectiveEqn}.  We note that $\overline{\hat{H}}_{\text{eff}}$ becomes time-independent only after time averaging.  This is in contrast to methods using cancellation of individual terms without time averaging \cite{Bukov2016}.

\section{Full Derivation of the Ising Interaction from the XXZ Hamiltonian} \label{app:fullDerivationIsing}

In this section we give the details allowing the derivation of an Ising interaction from an XXZ interaction using (global) pulse engineering.
We aim to construct pulses that accurately convert a 
general spin model, Eq.~\eqref{eq:spinModels}, into a desired form.  We choose to engineer $H_{ZZ}$ from $H_{XXZ}$ as an example. 

The problem is defined by inserting Eq.~\eqref{eq:spinModels}, Eqs.~\eqref{eq:Jsdipolar}, and Eq.~\eqref{eq:VasSigmaXandSigmaY} into Eq.~\eqref{eq:hamiltonianSystem}.  In this construction, 
we have $\alpha \in \{1, 2\}$ and $g_1$, $g_2$ are site-independent.  We can now use Eqs.~\eqref{eq:lowestKick} and ~\eqref{eq:effectiveEqn} to obtain an effective model to lowest order in driving inverse frequency:
\begin{align}
        \overline{\hat{H}^{\text{S},(0)}_{\text{eff}}} &= \sum_{p=0}^{\infty} \frac{(-1)^pv^{2p}}{(2p)!} \overline{G_1^{2p}}\left[\sum_{k_1} \hat{\sigma}^{(1)}_{k_1},...\left[\sum_{k_{2p}}\hat{\sigma}^{(1)}_{k_{2p}}, \hat{H}_0^{\text{S}}\right]...\right] \nonumber \\
        &+ \sum_{p=0}^{\infty} \frac{(-1)^pv^{2p}}{(2p)!} \overline{G_2^{2p}}\left[\sum_{k_1} \hat{\sigma}^{(2)}_{k_1},...\left[\sum_{k_{2p}}\hat{\sigma}^{(2)}_{k_{2p}}, \hat{H}_0^{\text{S}}\right]...\right]. \nonumber
    \end{align}
Working out the commutators requires analyzing four cases:
\begin{align}
    D^{2p}_1[\hat{H}_{XY}] &:= \sum_{k_1} \cdots \sum_{k_{2p}} \left[ \hat{\sigma}^{(1)}_{k_1},...\left[\hat{\sigma}^{(1)}_{k_{2p}}, \hat{H}_{XY}\right]...\right]
     \end{align}
     \begin{align}
    D^{2p}_1[\hat{H}_{ZZ}] &:= \sum_{k_1} \cdots \sum_{k_{2p}}  \left[\hat{\sigma}^{(1)}_{k_1},...\left[\hat{\sigma}^{(1)}_{k_{2p}}, \hat{H}_{ZZ}\right]...\right]
    \end{align}
    \begin{align}
    D^{2p}_2[\hat{H}_{XY}] &:= \sum_{k_1} \cdots \sum_{k_{2p}}  \left[ \hat{\sigma}^{(2)}_{k_1},...\left[\hat{\sigma}^{(2)}_{k_{2p}}, \hat{H}_{XY}\right]...\right]
    \end{align}
    \begin{align}
    D^{2p}_2[\hat{H}_{ZZ}] &:= \sum_{k_1} \cdots \sum_{k_{2p}}  \left[\hat{\sigma}^{(2)}_{k_1},...\left[\hat{\sigma}^{(2)}_{k_{2p}}, \hat{H}_{ZZ}\right]...\right].
\end{align}
Starting with the case $p = 1$, we see that $\hat{H}_0^{\text{S}}$ is `harmonic' in the sense that it satisfies $D^2[\hat{H}_0^{\text{S}}] := D^2_1[\hat{H}_0^{\text{S}}] + D^2_2[\hat{H}_0^{\text{S}}] \propto \hat{H}_0^{\text{S}}$. We then have, for example: 
\begin{align}
    D^2_1[\hat{H}_{XY}] &= \frac{J_{\perp}}{2}\sum_{j \neq j'} \mathcal{V}_{j,j'}
\bigg( \sum_{k, k'} D^1_{k, k'} [\hat{\sigma}^{(1)}_{j}\hat{\sigma}^{(1)}_{j'}]  \nonumber \\ 
&+ \sum_{k, k'} D^1_{k, k'}[\hat{\sigma}^{(2)}_{j}\hat{\sigma}^{(2)}_{j'}]
    \bigg), \nonumber
\end{align}
with $D^{\alpha}_{j,k}[-] = [\hat{\sigma}^{(\alpha)}_j,[\hat{\sigma}^{(\alpha)}_k,-]]$. If one expands these and simplifies, we obtain:
\begin{widetext}
\begin{align}
            D^2[\hat{H}_0^{\text{S}}] &= \frac{J_{\perp}}{2} \sum_{j \neq j'} 
            \mathcal{V}_{j,j'}
            \sum_{k, k'}\left(D^1_{k, k'}[\hat{\sigma}^{(2)}_{j} \hat{\sigma}^{(2)}_{j'}] + D^2_{k, k'}[\hat{\sigma}^{(1)}_{j} \hat{\sigma}^{(1)}_{j'}]\right)
            + \frac{J_z}{2} \sum_{j \neq j'}
            \mathcal{V}_{j,j'}
            \sum_{k, k'}\left(D^1_{k, k'}[\hat{\sigma}^{(3)}_{j} \hat{\sigma}^{(3)}_{j'}] + D^2_{k, k'}[\hat{\sigma}^{(3)}_{j} \hat{\sigma}^{(3)}_{j'}]\right).
   \end{align}
\end{widetext}
We find the general identity:
\begin{equation}
    D^2_{\alpha}[\hat{\sigma}^{(\beta)}_{j} \hat{\sigma}^{(\beta)}_{j'}] = 8(\hat{\sigma}^{(\beta)}_{j}\hat{\sigma}^{(\beta)}_{j'} - |\varepsilon^{\alpha\beta\gamma}|\hat{\sigma}^{(\gamma)}_{j} \hat{\sigma}^{(\gamma)}_{j'}),
    \label{eq:sigmaIdentity}
\end{equation}
which we use to establish a recursion relation. To this end, we use this identity in the above and simplify to obtain:
\begin{align}
    D_1^2[\hat{H}_0^{\text{S}}] &= 4\delta J \sum_{j \neq j'} 
     \mathcal{V}_{j,j'}
    \left(
    \hat{\sigma}^{(3)}_{j} \hat{\sigma}^{(3)}_{j'} - \hat{\sigma}^{(2)}_{j} \hat{\sigma}^{(2)}_{j'}
    \right)
    \\
    D^2_2[\hat{H}_0^{\text{S}}] &= 4\delta J \sum_{j \neq j'} 
    \mathcal{V}_{j,j'}
    \left(
    \hat{\sigma}^{(3)}_{j} \hat{\sigma}^{(3)}_{j'} - \hat{\sigma}^{(1)}_{j} \hat{\sigma}^{(1)}_{j'}
     \right).
\end{align}
Then we have the recursive calculations:
\begin{equation}
    D^4_{\alpha}[\hat{H}_0^{\text{S}}] = D^2_{\alpha}[D^2_{\alpha}[\hat{H}_0^{\text{S}}]], 
\end{equation}
from which we can obtain [by a second application of the identities Eq.~\eqref{eq:sigmaIdentity}]: 
$
    D^4_{\alpha}[\hat{H}_0^{\text{S}}] = 2\cdot8D^2_{\alpha}[\hat{H}_0^{\text{S}}],
$
which generalizes to:
\begin{equation}
    \begin{matrix}
        D^{2p}_{\alpha}[\hat{H}_0^{\text{S}}] = (2\cdot8)^{p-1}D^2_{\alpha}[\hat{H}_0^{\text{S}}], & p \geq 1.
    \end{matrix}
\end{equation}
After substitution into the expansion with the effective Hamiltonian, we find:
\begin{align}
        \overline{\hat{H}_{\text{eff}}^{\text{S},(0)}} &= \hat{H}_0^{\text{S}} + \frac{1}{16}\left(\sum_{p=1}^{\infty} \frac{(-1)^p(4v)^{2p}}{(2p)!}\overline{G^{2p}_1}\right)D^2_1[\hat{H}_0^{\text{S}}] \nonumber \\
        &+ \frac{1}{16}\left(\sum_{p=1}^{\infty} \frac{(-1)^p(4v)^{2p}}{(2p)!}\overline{G^{2p}_2}\right)D^2_2[\hat{H}_0^{\text{S}}].
    \end{align}
Defining:
\begin{equation}
    U^{(\alpha)} = \frac{1}{16}\sum_{p=1}^{\infty} \frac{(-1)^p(4v)^{2p}}{(2p)!}\overline{G^{2p}_{\alpha}},
\end{equation}
allows us to simplify:
\begin{widetext}
    \begin{align}
       \overline{\hat{H}^{\text{S},(0)}_{\text{eff}}} = \sum_{j \neq j'} \mathcal{V}_{j,j'}
        \left[
       (J_{\perp}/2 - 4\delta J U^{(2)})\hat{\sigma}^{(1)}_{j} \hat{\sigma}^{(1)}_{j'} + (J_{\perp}/2 - 4\delta J U^{(1)})\hat{\sigma}^{(2)}_{j} \hat{\sigma}^{(2)}_{j'}
        \right]
        + \sum_{j \neq j'}
        \mathcal{V}_{j,j'}
        \left[\frac{J_z}{2} + 4 \delta J (U^{(1)} + U^{(2)})\right] \hat{\sigma}^{(3)}_{j} \hat{\sigma}^{(3)}_{j'}
        \label{eq:effH0XXZ}.
    \end{align}
\end{widetext}
$\overline{\hat{H}^{\text{S},(0)}_{\text{eff}}}$ allows us to use Floquet engineering to inter-relate a large class of starting Hamiltonians and effective Hamiltonians.  For example, 
if we want to obtain $\overline{\hat{H}^{(0)}_{\text{eff}}} \propto \hat{H}_{ZZ}$, we need:
\begin{align}
    U^{(1)} = U^{(2)} = \frac{J_{\perp}}{8\delta J},
\end{align}
or equivalently:
\begin{equation}
    \sum_{p=1}^{\infty}\frac{(-1)^p(4v)^{2p}}{(2p)!}\overline{G^{2p}_{\alpha}} = \frac{2}{s-1},
\end{equation}
 where $s = J_z/J_{\perp}$. Notice that this is satisfied if $G_1 = G_2 = G$ so that: 
\begin{equation}
    \overline{\hat{H}^{\text{S},(0)}_{\text{eff}}} = \frac{J_z + 2J_{\perp}}{2}\sum_{j \neq j'} 
    \mathcal{V}_{j,j'}
    \hat{\sigma}^{(3)}_{j} \hat{\sigma}^{(3)}_{j'}.
\end{equation}
We have therefore shown that pulse engineering can be used to convert the XXZ model to an effective Ising model at lowest order in inverse frequency of the driving pulse.  

\section{Example Physical Systems}
\label{app:physicalsystems}

We briefly review how Alkali-based polar molecules, e.g., CsRb, and Rydberg atoms lead to effective spin models.  Such experiments are captured by  \cite{DeMille2002,Barnett2006,WALL2010a,Saffman2010,muller2010,Wall2014,NGUYEN2018} Eqs.~\eqref{eq:hamiltonianSystem},  \eqref{eq:spinModels}, and \eqref{eq:Jsdipolar} because in both cases, particles can be trapped in optical tweezers or optical lattices \cite{NGUYEN2018,BLUVSTEIN2021a,EBADI2021b,scholl2022microwave,CHRISTAKIS2023b,Holland2023,Bao2023}.  Localization of the particles into well-defined sites and mapping to pseudospin degrees of freedom implies that these systems are captured by spin lattice models.  

Polar molecules in certain rotational states interact due to their dipole moment, $d_0$.  Modeling the molecules as dipole moments in a planar array, e.g., a square lattice, we can define a quantization axis along a symmetry-breaking constant (DC) electric field pointing perpendicular to the plane.  To define a pseudospin one can restrict to the subspace of rotational states where $\mathcal{N} \in \{0,1\}$ is the total rotational angular momentum
and $m_{\mathcal{N}} = 0$ is the $z$-component of the angular momentum \cite{DeMille2002}. 

A dipolar XXZ model arises due to Stark shift in the energy levels caused by the electric field. In that case, $|\uparrow\rangle$ is the Stark-shifted $|\mathcal{N}=0,m_\mathcal{N} = 0\rangle$ state and $|\downarrow\rangle$ the Stark-shifted $|\mathcal{N} = 1, m_{\mathcal{N}} = 0\rangle$ state. In so doing, we obtain values for $J_{\perp}$ and $J_z$ that depend on the transition amplitudes of the Stark-shifted states and $d_0$.  One can then show that Eq.~\eqref{eq:spinModels} offers an excellent approximation where \cite{Barnett2006,muller2010}:
\begin{align}
    J_{\perp} &= \frac{(\langle\uparrow|d_0|\downarrow\rangle)^2}{a^3}\\
    J_z &= \frac{(\langle \uparrow|d_0|\uparrow\rangle - \langle \downarrow|d_0|\downarrow\rangle)^2}{2a^3},
\end{align}
$a$ is the distance between nearest neighbors in the two-dimensional lattice, and the site-dependence of the interaction has a dipolar form, $\mathcal{V}_{j,j'}=\vert j-j'\vert^{-3}$.  The relative interaction strengths can be characterized by an angle $\theta = \tan^{-1}(J_{\perp}/J_z)$ that is tunable with the external electric field strength.  The domain of $\theta$ for these physical systems ranges from $\pi/2$ to about $\pi/9$ so that the relative strengths cannot be negative \cite{muller2010}.  This shows that a simple definition of a pseudospin using just rotational states of Alkali-based polar molecules leads to a sign restriction in $J_{\perp}$ and $J_z$.  The restriction in turn limits the ability for global pulse protocols to convert a dipolar XXZ model into a purely Ising model.   It would be interesting to explore the possibility of interleaving pulse schemes \cite{Lee2016} designed to eliminate the long-range tail of the dipolar interaction to engineer bipartite models for molecules. 

Recent experiments with Rydberg atoms have been able to engineer XY models and study their tunability \cite{GEIER2021c,scholl2022microwave,ZHAO2023}. 
Work studying circular Rydberg atoms trapped in optical tweezer arrays \cite{NGUYEN2018} also argues for the ability to realize 
dipolar XXZ models where a relative minus sign between the XY and ZZ terms can be obtained \cite{NGUYEN2018}. Consequently, circular Rydberg atom systems can be converted from a dipolar XXZ model to a purely Ising model using global pulsing. 

\bibliography{Refs}% Produces the bibliography via BibTeX.

\end{document}